\documentclass[11pt, a4paper]{article}
\pdfoutput=1
\usepackage[utf8]{inputenc}
\usepackage[T1]{fontenc}
\usepackage{lmodern}
\usepackage{textcomp}
\usepackage[american]{babel}
\usepackage{microtype}
\usepackage{csquotes}

\usepackage{eurosym}
\usepackage{xspace}
\usepackage[nottoc]{tocbibind}
\usepackage{authblk}
\usepackage{fancyhdr}
\usepackage{setspace}
\usepackage{titlesec}
\usepackage{changepage}
\usepackage{totpages}

\usepackage{graphicx}
\usepackage{subcaption}
\usepackage{float}
\usepackage{makecell}
\usepackage{multirow}
\usepackage{multicol}
\usepackage{enumitem}
\usepackage{marginnote}
\usepackage[multiple]{footmisc}

\usepackage[usenames, dvipsnames, svgnames, table]{xcolor}

\usepackage[
	backend=bibtex8, citestyle=numeric-comp,
	sorting=none, sortcase=false, sortcites=true,
	giveninits=true, maxnames=99
]{biblatex}

\usepackage{amsmath, amsfonts, amssymb}
\usepackage{mathtools}
\usepackage{mathrsfs}
\usepackage{cancel}
\usepackage{braket}
\usepackage{tensor}
\usepackage{slashed}
\usepackage{siunitx}

\usepackage{listings}

\usepackage{framed, mdframed}
\usepackage[amsmath, hyperref, framed]{ntheorem}

\usepackage[
	pdftex, breaklinks=true, linktocpage, colorlinks=true,
	urlcolor=RoyalBlue, linkcolor=RoyalBlue, citecolor=BrickRed
]{hyperref}
\usepackage[all]{hypcap}

\DeclareUnicodeCharacter{00A0}{~}
\DeclareUnicodeCharacter{202F}{~}

\newcommand{\email}[1]{\href{mailto:#1}{\nolinkurl{#1}}}
\newcommand{\emailfoot}[1]{\thanks{\email{#1}}}

\newcommand{\doi}[1]{\href{http://dx.doi.org/#1}{\nolinkurl{#1}}}
\newcommand{\urlhttp}[1]{\href{http://#1}{\nolinkurl{#1}}}

\newcounter{draftcommentcnt}
\NewDocumentCommand{\draftcomment}{s O{red} m}{%
	\def\margnote{\IfBooleanTF{#1}{\marginnote}{\marginpar}}%
	\stepcounter{draftcommentcnt}%
	\textcolor{#2}{#3}%
	\margnote{\textcolor{#2}{$\Leftarrow$ \arabic{draftcommentcnt}}}%
}

\graphicspath{{./images/}}
\numberwithin{equation}{section}

\usepackage[sort&compress, english]{cleveref}

\usepackage[heightrounded,
	top=3.5cm, bottom=3cm, left=3.5cm, right=3.5cm,
]{geometry}

\fancypagestyle{pageof}{
	\fancyhf{}

	\fancyfoot[C]{-- Page \thepage\ of \pageref*{TotPages} --}
}

\newcommand{\preprint}[0]{XXX}
\fancypagestyle{preprint}{
	\fancyhf{}

	\fancyhead[R]{\small\textsf{\preprint}}
}

\lstdefinestyle{boxed}{frame=single, numbers=left}

\newcommand{\I}[0]{\mathrm{i}}

\newcommand{\Z}[0]{\mathbb{Z}}

\newcommand{\C}[0]{\mathbb{C}}

\newcommand{\dd}[0]{\mathrm{d}}

\newcommand{\mc}[1]{{\mathcal{#1}}}

\renewcommand{\bra}[2][]{\mathinner{{}_{#1}\hspace{-2pt}\langle #2 |}}

\newcommand{\acom}[1]{\{ #1 \}}

\newcommand{\conj}[1]{{#1^*}}
\newcommand{\adj}[1]{{#1^\dagger}}
\newcommand{\bpz}[1]{{#1^t}}

\input{arxiv.def}

\hypersetup{
	pdftitle={Closed string theory without level-matching at the free level},
	pdfauthor={Harold Erbin}
}

\title{Closed string theory without level-matching at the free level}

\author[1,2,3]{Harold Erbin\emailfoot{erbin@mit.edu}}
\author[4]{Maxime Médevielle\emailfoot{maxime.medevielle@liverpool.ac.uk}}

\affil[1]{%
	Center for Theoretical Physics, Massachusetts Institute of Technology
	\protect\\
	Cambridge, MA 02139, \textsc{Usa}
}

\affil[2]{%
	\textsc{Nsf Ai} Institute for Artificial Intelligence and Fundamental Interactions
}

\affil[3]{%
	Université Paris Saclay, \textsc{Cea}, \textsc{List}, Gif-sur-Yvette, F-91191, France
}

\affil[4]{%
	Department of Mathematical Sciences, University of Liverpool
	\protect\\Peach Street, Liverpool L69 7\textsc{zl}, \textsc{Uk}
}

\allowdisplaybreaks

\begin{document}

\maketitle

\begin{abstract}
In its traditional form, the string field in closed string field theory is constrained by the level-matching condition, which is imposed beside the action.
By analogy with the similar problem for the Ramond sector, it was understood by Okawa and Sakaguchi how to lift this condition and work with unconstrained field by introducing spurious free fields.
These authors also pointed out that new backgrounds may exist thanks to a new gauge field which is trivial on flat space, but can generate fluxes on a toroidal background.
In this paper, we perform a complete study of the free theory at the tachyonic and massless levels with the aim of setting the stage for studying backgrounds without level-matching.
\end{abstract}

\vspace{\stretch{1}}
\newpage

\hrule
\pdfbookmark[1]{\contentsname}{toc}
\tableofcontents
\bigskip
\hrule

\newpage

\section{Introduction}

Physical closed string states satisfy the level-matching condition~\cite{Nelson:1989:CovariantInsertionGeneral,La:1990:EffectiveFieldEquations,Distler:1991:TopologicalCouplingsContact,Becchi:1994:SemirelativeConditionClosed,Erbin:2021:StringFieldTheory}
\begin{subequations}
\label{eq:level-matching}
\begin{equation}
    L_0^-
        := L_0 - \bar L_0
        = 0,
\end{equation}
where $L_0^-$ generates translation along the periodic spatial direction $\sigma$ of the string using cylinder coordinates.
This is necessary to fix the origin of $\sigma$ when parametrizing the closed string worldsheet: indeed, there is no natural gauge fixing condition and this must be imposed as a constraint.
In the BRST formalism, one needs to impose the equivalent condition on the associated $b$-ghost:
\begin{equation}
    b_0^-
        := b_0 - \bar b_0
        = 0.
\end{equation}
\end{subequations}
This condition extends to off-shell states by investigating the properties of off-shell amplitudes: metrics in patches whose coordinates differ by a phase cannot be distinguished (rotations in the complex plane are generated by $L_0^-$).
A corollary is that the antibracket from the BV algebra on the moduli spaces of Riemann surfaces with general local coordinates is degenerate~\cite{Sen:1996:BackgroundIndependentAlgebraic}.
Moreover, moduli space topology implies that no global choice of local coordinates exists, making off-shell amplitudes without level-matching multi-valued.
Finally, \eqref{eq:level-matching} becomes clearly apparent in string field theory (SFT)~\cite{Zwiebach:1993:ClosedStringField,Hull:2009:DoubleFieldTheory,deLacroix:2017:ClosedSuperstringField,Erbin:2021:StringFieldTheory}: the propagator obtained by factorizing amplitudes in Siegel gauge contains a factor $b_0^- \delta(L_0^-)$.
Inverting the propagator gives a degenerate kinetic term if the string field does not obey \eqref{eq:level-matching} and, while the degeneracy due to the Siegel gauge condition $b_0^+ := b_0 + \bar b_0 = 0$ can be lifted by restoring the gauge invariance, this does not appear to be possible for the level-matching condition.
Similarly, the natural inner-product between closed string states (used for example to define string products from interaction vertices) includes a factor $c_0^- \delta(L_0^-)$.

This situation is not completely satisfactory.
Indeed, fields in a fundamental theory should arguably not be restricted by any condition outside representation theory, and all constraints should derive from dynamics or gauge fixing.
Additionally, constraints can be understood as degrees of freedom integrated out, which makes the theory more complicated than it should be~\cite[p.~28]{Erler:2020:FourLecturesClosed}.
Recently, it was understood by Okawa and Sakaguchi~\cite{Okawa:2022:ClosedStringField,Okawa:2018:ClosedStringField-Talk} that the level-matching conditions \eqref{eq:level-matching} can be lifted by introducing a new spurious field (see also~\cite[p.~10]{Moosavian:2020:SuperstringFieldTheory}).
This mirrors exactly what happens when attempting to describe the Ramond sector in the superstring theory.
While the kinetic term for the NS sector is not degenerate (after imposing the level-matching condition for the closed superstring), it is not the case for the Ramond sector.
One solution is to impose constraints on the field and define a non-degenerate inner-product~\cite{Kunitomo:2016:CompleteActionOpen,Erler:2016:CompleteActionOpen,Erler:2020:FourLecturesClosed}, which is formally the same as imposing the level-matching conditions.
Another approach was proposed in~\cite{Sen:2015:GaugeInvariant1PI-R,Sen:2015:CovariantActionType,Sen:2016:BVMasterAction,deLacroix:2017:ClosedSuperstringField}, where it was shown that a free field could be introduced to write an unconstrained Ramond field.
This new spurious field has a non-trivial kinetic term and seems to couple to the original string field, but it can be shown that it describes only free degrees of freedom.
Moreover, the additional degrees of freedom in the unconstrained string field are not physical, and correspond to gauge and auxiliary fields.
Given the analogy mentioned above, the same strategy can be followed to relax \eqref{eq:level-matching}~\cite{Okawa:2022:ClosedStringField,Okawa:2018:ClosedStringField-Talk}.

Removing the level-matching condition is interesting for several reasons.
First, it was observed in~\cite{Okawa:2022:ClosedStringField,Okawa:2018:ClosedStringField-Talk} that the additional fields, which are trivial in a flat background, can yield fluxes in a toroidal background.
Hence, they can lead to new backgrounds whose physics should be explored, for example, its $T$-duality properties.
Second, the level-matching condition descends to the weak constraint in double field theory~\cite{Hull:2009:DoubleFieldTheory}, such that the same technique could be relevant to formulate double field theory without constraint.
Finally, interacting closed SFT is non-polynomial and difficult to describe: removing the level-matching condition could help in finding new simpler formulations~\cite{Erler:2020:FourLecturesClosed}, ideally cubic.
Indeed, this new formulation allows for more flexibility in writing the kinetic term as the operator $B$ can be replaced by other operators (again, in full analogy with the superstring, where different kinetic terms are used)~\cite{Okawa:2022:ClosedStringField}.

In this paper, we focus on free closed bosonic SFT.
Our goal is to understand better this theory after removing the level-matching condition, with the long-term goal of studying $T$-duality and Buscher rules.
We will expand the string field in terms of spacetime fields up to the massless sectors, compute the equations of motion, action and gauge transformations.
The main additions of our work to~\cite{Okawa:2022:ClosedStringField,Okawa:2018:ClosedStringField-Talk} are: 1) working out the massless even sector (containing the $B$ field, which is needed to discuss $T$-duality), and the level $(2, 0)$, then 2) showing that fields are non-dynamical by using field redefinitions to reach a formulation transforming canonically under field redefinitions, 3) making explicit the reality condition on the new fields and gauge parameters, proving that the action is real.

In \Cref{sec:sft}, we will set up our notations and reproduce the derivation from~\cite{Okawa:2022:ClosedStringField,Okawa:2018:ClosedStringField-Talk} of the SFT action without level-matching.
In \Cref{sec:level-exp}, we compute the spacetime properties for the levels $(0, 0)$ (tachyon),  $(1, 0)$ (non-level-matched tachyon),  $(1, 1)$ (massless),  $(2, 0)$ (non-level matched massless).

\medskip

\emph{
Note: The current paper is scheduled to appear together with~\cite{Okawa:2022:ClosedStringField}.
A large part of this work has been realized independently, using the results announced in~\cite{Okawa:2018:ClosedStringField-Talk}.
However, the authors from~\cite{Okawa:2022:ClosedStringField} have shared their draft such that we could compare our results with theirs.
We tried to provide reference to~\cite{Okawa:2022:ClosedStringField} for every result derived there, and we encourage the readers to read~\cite{Okawa:2022:ClosedStringField} first, as it provides more details on the construction, in particular, including interactions.
}

\section{Closed string field theory without level-matching}
\label{sec:sft}

In this section, we start by reviewing the free string field theory (SFT) for the closed bosonic string (\Cref{sec:sft:level-matched}).
In particular, we show how the level-matching condition \eqref{eq:level-matching} appears to be necessary to write the kinetic term.
Afterwards, we will show how this condition can be removed by extending the space in which lives the string field (\Cref{sec:sft:unconstrained}).

\subsection{Closed string field theory}
\label{sec:sft:level-matched}

String field theory is a second quantized version of worldsheet string theory.
The string background on which the theory lives is described by a $2d$ theory made of a matter sector and a $bc$ reparametrization ghost sector.
In this paper, we consider a flat Minkowski background with $d = 26$ free scalar fields $X^\mu$ ($\mu = 0, \ldots, d-1$) for the matter sector.
We provide the necessary CFT definitions and formulas in \Cref{app:formulas}, and refer the reader to the literature for more details~\cite{Zwiebach:1993:ClosedStringField,deLacroix:2017:ClosedSuperstringField,Erler:2020:FourLecturesClosed,Erbin:2021:StringFieldTheory}.

The string field is a general element of the CFT Hilbert space restricted to states with ghost number $2$ satisfying the level-matching condition \eqref{eq:level-matching}:
\begin{equation}
    N_{\text{gh}}(\Psi)
        = 2,
    \qquad
    L_0^- \ket{\Psi}
        = b_0^- \ket{\Psi}
        = 0,
\end{equation}
together with the reality condition
\begin{equation}
    \label{eq:sft:reality}
    \adj{\ket{\Psi}}
        = - \bra{\Psi},
\end{equation}
where $\bra{\Psi} := \bpz{\ket{\Psi}}$ is the BPZ conjugate of $\ket{\Psi}$ and the dagger denotes Euclidean conjugation (both defined in \Cref{app:formulas}).

The classical action for the free closed bosonic SFT is:
\begin{equation}
    \label{eq:sft:action}
	S[\Psi]
        = \frac{1}{2} \, \bra{\Psi} \delta(L_0^-) c_0^- Q_B \ket{\Psi},
\end{equation}
where $Q_B$ is the BRST charge.
This action is real thanks to \eqref{eq:sft:reality}.
The $c_0^-$ insertion is necessary to reach $N_{\text{gh}} = 6$, as required by the ghost number anomaly on the sphere.
Hence, the natural inner-product in closed string theory is $\bra{\cdot} c_0^- \ket{\cdot}$.

The action \eqref{eq:sft:action} clearly shows why it is necessary to impose the level-matching condition \eqref{eq:level-matching}.
Indeed, an unconstrained field can be decomposed as:
\begin{equation}
    \ket{\Psi}
        = \ket{\Psi_-} + c_0^- \ket{\Psi_+},
    \qquad
    b_0^- \ket{\Psi_\pm}
        = 0.
\end{equation}
Inserting such a field in the action yields a degenerate kinetic term, since $\bra{\Psi} c_0^- = \bra{\Psi_-} c_0^-$.
The constraint $b_0^- \ket{\Psi} = 0$ implies that $\Psi_+ = 0$, and the kinetic term is well-defined.

The action is invariant under the gauge transformation
\begin{equation}
    \label{eq:sft:gauge-inv}
    \delta_{\Lambda} \ket{\Psi}
        = Q_B \ket{\Lambda},
\end{equation}
where the gauge parameter satisfies the condition:
\begin{equation}
    N_{\text{gh}}(\Lambda)
        = 1,
    \qquad
    L_0^- \ket{\Lambda}
        = b_0^- \ket{\Lambda}
        = 0.
\end{equation}
Physical states are solutions to the equation of motion
\begin{equation}
    \label{eq:sft:eom}
    Q_B \ket{\Psi}
        = 0,
\end{equation}
up to gauge transformations, since $c_0^-$ is invertible in the subspace of level-matched states.

Not all $\Lambda$ gauge parameters are independent, since the gauge transformation is trivial when $\ket{\Lambda} = Q_B \ket{\Omega}$.
Indeed, $Q_B^2 = 0$ implies that $\Lambda$ itself transforms as:
\begin{equation}
    \label{eq:sft:gauge-transf-Lambda}
    \delta_{\Omega} \ket{\Lambda}
        = Q_B \ket{\Omega},
    \qquad
    N_{\text{gh}}(\Omega)
        = 0,
    \qquad
    L_0^- \ket{\Omega}
        = b_0^- \ket{\Omega}
        = 0.
\end{equation}
In turn, there is a gauge invariance of $\Omega$, and this continues recusively: however, we will not need to consider more parameters in this paper.

The gauge symmetry \eqref{eq:sft:gauge-inv} can be fixed with the Siegel condition:
\begin{equation}
    b_0^+ \ket{\Psi}
        = 0.
\end{equation}
In this case, the action \eqref{eq:sft:action} becomes
\begin{equation}
    \label{eq:sft:action-gf}
	S_{\text{gf}}[\Psi]
        = \frac{1}{2} \, \bra{\Psi} \delta(L_0^-) c_0^- c_0^+ L_0^+ \ket{\Psi},
\end{equation}
which allows identifying the propagator:
\begin{equation}
    \Delta
        := \frac{b_0^+}{L_0^+} \, B,
    \qquad
    B
        := b_0^- \delta(L_0^-).
\end{equation}
The string propagator is a cylinder with two moduli: a length (proper-time) $s$ and a twist $\theta$.
Scale dilatations correspond to shifting $s$ and are generated by $L_0^+$, rotations are generated by $L_0^-$; $b_0^+$ and $b_0^-$ are the BRST anti-ghost modes corresponding to these transformations (in off-shell string theory, they arise from Beltrami differentials used to write a form on the moduli space).
In momentum space, the first term in the RHS becomes the well-known $(k^2 + m^2)^{-1}$ from local QFT (see \Cref{sec:level-exp}), while the second term $B$ has no direct QFT interpretation.

Since $(B c_0^-)^2 = B c_0^-$, the operator $B$ can be used to rewrite the level-matching condition \eqref{eq:level-matching} as a projection equation~\cite{Okawa:2022:ClosedStringField}:
\begin{equation}
    B c_0^- \ket{\Psi}
        = \ket{\Psi}.
\end{equation}
This is formally the same as the condition $X Y \ket{\Psi} = \ket{\Psi}$ for the Ramond sector superstring field ($X$ and $Y$ are so-called picture changing operators and are not needed in this paper)~\cite{Kunitomo:2016:CompleteActionOpen,Erler:2016:CompleteActionOpen,Erler:2020:FourLecturesClosed}.
Note also that operator $B$ commutes with the BRST operator:
\begin{equation}
	\acom{Q_B, B}
		= \delta(L_0^-) \, \acom{Q_B, b_0^-}
		= \delta(L_0^-) \, L_0^-
		= 0.
\end{equation}

\subsection{Removing the level-matching condition}
\label{sec:sft:unconstrained}

Given an operator $K$ on a Hilbert space $\mc H$ which is not invertible, there are two ways to define a non-degenerate inner-product.
The first, as we have seen above, is to restrict the Hilbert space to the subspace where $K$ is invertible.
The second runs in the opposite direction and consists in embedding $\mc H$ into a higher-dimensional space $\mc H_{\text{ext}}$ and defining a new operator $A + B K$ (with $A$ and $B$ matrices) which is invertible in this space, without using $K^{-1}$.
This idea was successfully applied to the Ramond sector in~\cite{Sen:2015:GaugeInvariant1PI-R,Sen:2015:CovariantActionType,Sen:2016:BVMasterAction}, and it was understood in~\cite{Okawa:2022:ClosedStringField,Okawa:2018:ClosedStringField-Talk} that it can also be used for the level-matching condition.
In the rest of this subsection, we summarize the results from~\cite{Okawa:2022:ClosedStringField,Okawa:2018:ClosedStringField-Talk} relevant for the free theory, and refer the readers to them for more details.

Starting from the gauge-fixed action \eqref{eq:sft:action-gf}, we introduce a new spurious\footnotemark{} field $\widetilde{\Psi}$ and write
\footnotetext{%
    This is not an auxiliary field because it is dynamical, whereas an auxiliary field would have an algebraic equation of motion.
    However, its main role is to behave like a Lagrange multiplier in the action, without affecting the dynamics since it remains free.
}%
\begin{equation}
    \label{eq:sft:action-ext-gf}
    S_{\text{ext,gf}}[\Psi, \widetilde{\Psi}]
        = \frac{1}{2} \, \bra{\widetilde{\Psi}} c_0^+ L_0^+ B \ket{\widetilde{\Psi}}
		    + \bra{\widetilde{\Psi}} c_0^+ L_0^+ \ket{\Psi},
\end{equation}
where we need
\begin{equation}
    N_{\text{gh}}(\widetilde{\Psi})
        = 3
\end{equation}
from the ghost anomaly on the sphere.
The kinetic operator in the basis $(\widetilde{\Psi}, \Psi)$ reads
\begin{equation}
	K
        := c_0^+ L_0^+
            \begin{pmatrix}
                B & 1 \\
                1 & 0
            \end{pmatrix},
\end{equation}
and its inverse is:
\begin{equation}
	K^{-1} = \frac{b_0^+}{L_0^+}
		\begin{pmatrix}
			0 & 1 \\
			1 & B
		\end{pmatrix}.
\end{equation}
The lower-right component reproduces the propagator $\Delta$ between string fields $(\Psi, \Psi)$.
Given this structure, it is clear that states which do not satisfy \eqref{eq:level-matching} have a non-trivial propagator, and the inner-product in the action is not degenerate.

It is straightforward to remove the Siegel gauge fixing condition to get the gauge invariant action:
\begin{equation}
    \label{eq:sft:action-ext}
	S_{\text{ext}}[\Psi, \widetilde{\Psi}]
        = \frac{1}{2} \, \bra{\widetilde{\Psi}} Q_B B \ket{\widetilde{\Psi}}
            + \bra{\widetilde{\Psi}} Q_B \ket{\Psi}.
\end{equation}
Both fields $\Psi$ and $\widetilde{\Psi}$ are completely unconstrained (except for the ghost number condition).
The action is invariant under two sets of transformations:
\begin{subequations}
\begin{gather}
    \delta_{\Lambda} \ket{\Psi}
        = Q_B \ket{\Lambda},
    \qquad
    \delta_{\Lambda} \ket{\widetilde{\Psi}}
        = 0,
    \\
    \delta_{\tilde{\Lambda}} \ket{\Psi}
        = 0,
    \qquad
    \delta_{\tilde{\Lambda}} \ket{\widetilde{\Psi}}
        = Q_B \ket{\widetilde{\Lambda}},
\end{gather}
\end{subequations}
where the gauge parameters have no constraints, except:
\begin{equation}
    N_{\text{gh}}(\Lambda)
        = 1,
    \qquad
    N_{\text{gh}}(\widetilde{\Lambda})
        = 2.
\end{equation}
While we are not interested in interactions, let us note that they take the same form as in the constrained closed SFT, which means that $\widetilde{\Psi}$ only appears in the quadratic terms.
Like for \eqref{eq:sft:gauge-transf-Lambda}, not all $\Lambda$ and $\widetilde{\Lambda}$ gauge parameters are independent~\cite{Okawa:2022:ClosedStringField}:
\begin{equation}
    \label{eq:sft:gauge-inv-gauge-param}
    \delta_{\Omega} \ket{\Lambda}
        = Q_B \ket{\Omega},
    \qquad
    N_{\text{gh}}(\Omega)
        = 0,
    \qquad
    \delta_{\tilde \Omega} \ket{\widetilde{\Lambda}}
        = Q_B \ket{\widetilde{\Omega}},
    \qquad
    N_{\text{gh}}(\widetilde{\Omega})
        = 1.
\end{equation}
Like before, the second-order parameters have themselves a gauge invariance, and so on.

We can recover the original action \eqref{eq:sft:action} for level-matched components by imposing the following constraints:
\begin{equation}
    \label{eq:sft:field-aux-to-phys}
    \ket{\widetilde{\Psi}}
        = - \delta(L_0^-) \, c_0^- \ket{\Psi},
    \qquad
    L_0^- \ket{\Psi}
        = L_0^- \ket{\widetilde{\Psi}}
        = 0,
    \qquad
    b_0^- \ket{\Psi}
        = c_0^- \ket{\widetilde{\Psi}}
        = 0,
\end{equation}
where the sign in the first condition is a consequence of $c_0^-$ being BPZ odd.
Note that, even though the gauge symmetry is larger in the theory without level-matching (because of the additional parameters $\delta(L_0^-) \ket{\Lambda}$, $b_0^- \ket{\Lambda}$ and $\ket{\widetilde{\Lambda}}$), the original theory cannot be reached by gauge fixing because the field combinations above are generically not BRST closed.\footnotemark{}
\footnotetext{%
    We thank Ted Erler for discussions.
}%
Rather, they are analogue to self-duality constraints which must be imposed beside the action.
We want to stress that we are not setting to zero all the degrees of freedom in $\widetilde \Psi$, since this would be inconsistent with the non-linear equations of motion, but also remove the kinetic term in the action (free or interacting).
Instead, we set those free fields components equal to the ones of $\Psi$.
This is exemplified at level $(0, 0)$ in \Cref{sec:level-exp:00:action}.
Exactly the same holds for the superstring, where the original theory is recovered by setting $\widetilde \Psi = \Psi$ (NS sector) and $\widetilde \Psi = Y \Psi$ (R sector), which is equivalent to the condition $\mc G \widetilde \Psi = \Psi$~\cite{Sen:2015:GaugeInvariant1PI-R} for fields satisfying $\Psi = X Y \Psi$~\cite{Erler:2016:CompleteActionOpen} (where $X$ and $Y$ are PCO operators, $\mc G$ is the identity in the NS sector, and $X$ in the Ramond sector).

The action \eqref{eq:sft:action-ext} is real if:
\begin{equation}
    \label{eq:sft:reality-ext}
    \adj{\ket{\Psi}}
        = - \bra{\Psi},
    \qquad
    \adj{\ket{\widetilde{\Psi}}}
        = \bra{\widetilde{\Psi}}.
\end{equation}
Indeed, we have:
\begin{align*}
    \adj{\braket{\widetilde{\Psi}, Q_B B \widetilde{\Psi}}}
        &
        = \braket{B Q_B \adj{\widetilde{\Psi}}, \adj{\widetilde{\Psi}}}
        = \braket{B Q_B \widetilde{\Psi}, \widetilde{\Psi}}
        = - \braket{\widetilde{\Psi}, B Q_B \widetilde{\Psi}}
        = \braket{\widetilde{\Psi}, Q_B B \widetilde{\Psi}},
    \\
    \adj{\braket{\widetilde{\Psi}, Q_B \Psi}}
        &
        = \braket{Q_B \adj{\Psi}, \adj{\widetilde{\Psi}}}
        = - \braket{Q_B \Psi, \widetilde{\Psi}}
        = \braket{\widetilde{\Psi}, Q_B \Psi}.
\end{align*}
For the first equation, we took the Euclidean conjugate in the first step, using that $Q_B$ and $B$ are Euclidean self-adjoint.
Then, we inserted the reality conditions \eqref{eq:sft:reality-ext}, used the relation \eqref{eq:cft:bpz-exch} (there is a sign since $\widetilde{\Psi}$ and $B Q_B \widetilde{\Psi}$ are both Grassmann-odd), and finally that $B$ and $Q_B$ anti-commute with each other.
The second derivation is similar.
This can be expected since $N_{\text{gh}}(\widetilde{\Psi}) = 3$, which means that the Euclidean conjugate of $\widetilde{\Psi}$ contains one additional $c$ ghost compared to $\Psi$, giving an additional sign in the BPZ conjugate given \eqref{eq:cft:bpz-conj}.
This condition was not discussed in full generality in~\cite{Okawa:2022:ClosedStringField}, but agrees with the conditions on the level expansion, assuming that all spacetime fields are real.
Note that it is not the same condition as in~\cite{Sen:2016:RealitySuperstringField}, since the spurious fields for the Ramond sector have $N_{\text{gh}} = 2$.

The gauge parameters obey the following reality conditions:
\begin{equation}
    \label{eq:sft:reality-ext-gauge-params}
    \adj{\ket{\Lambda}}
        = - \bra{\Lambda},
    \qquad
    \adj{\ket{\widetilde{\Lambda}}}
        = - \bra{\widetilde{\Lambda}}.
\end{equation}
To obtain these relations, we impose that $Q_B \ket{\Lambda}$ and $Q_B \ket{\widetilde{\Lambda}}$ have the same transformation as $\Psi$ and $\widetilde{\Psi}$ respectively:
\begin{align*}
    \adj{(Q_B \ket{\Lambda})}
       &
       = \bra{\adj{\Lambda}} Q_B,
    &
    \bra{Q_B \Lambda}
        &
        = - \bra{\Lambda} Q_B^t
        = \bra{\Lambda} Q_B,
    \\
    \adj{(Q_B \ket{\widetilde{\Lambda}})}
        &
        = \bra{\adj{\widetilde{\Lambda}}} Q_B,
    &
    \bra{Q_B \widetilde{\Lambda}}
        &
        = \bra{\widetilde{\Lambda}} Q_B^t
        = - \bra{\widetilde{\Lambda}} Q_B.
\end{align*}
For $\Lambda$, we used that $Q_B$ is Euclidean self-adjoint (first equation), and that it is BPZ odd and anti-commutes with $\Lambda$ since $N_{\text{gh}}(\Lambda) = 1$ (second equation).
Matching with \eqref{eq:sft:reality-ext} implies the relation above for $\Lambda$.
The only difference with $\widetilde{\Lambda}$ is that $N_{\text{gh}}(\widetilde{\Lambda}) = 2$, such that its state commutes with $Q_B$, and we need a different sign for the reality condition.

The equations of motion obtained by varying \eqref{eq:sft:action-ext} with respect to $\widetilde{\Psi}$ and $\Psi$:
\begin{subequations}
\begin{gather}
    Q_B B \ket{\widetilde{\Psi}} + Q_B \ket{\Psi}
        = 0,
    \\
    Q_B \ket{\widetilde{\Psi}}
        = \ket{J(\Psi)}.
\end{gather}
\end{subequations}
We have added a $\Psi$-dependent source (which would get replaced by interactions in the full theory) for illustrating why $\widetilde{\Psi}$ contains only free degrees of freedom.
Multiplying the second equation by $B$ and using the first equation to substitute $\widetilde{\Psi}$, we get (remember that $Q_B$ anti-commutes with $B$)
\begin{equation}
    Q_B \ket{\Psi}
        = B \ket{J(\Psi)}.
\end{equation}
This reproduces the equation of motion \eqref{eq:sft:eom} for the field $\Psi$ with a source $B \ket{J(\Psi)}$ when $\Psi$ satisfies \eqref{eq:level-matching}.
For the full proof of equivalence in the presence of a source, we refer the reader to~\cite{Okawa:2022:ClosedStringField}.
Once a solution for $\Psi$ is found from the previous equation, the second equation of motion fixes $\widetilde{\Psi}$ up to a free field.
As a consequence, both theories with and without level-matching are perturbatively equivalent.

\section{Tachyonic and massless spacetime actions}
\label{sec:level-exp}

The objective of this section is to provide an extensive analysis of tachyonic and massless sectors of the free closed SFT.
To achieve this, we perform a level expansion of the string field, impose reality conditions, and compute the action \eqref{eq:sft:action-ext} together with the equations of motion and gauge transformations.

Next, in order to show that the additional spacetime fields are not physical, we perform field redefinitions to work with fields which have canonical gauge transformations.
Indeed, it is well-known that the spacetime fields appearing in the string field expansion have non-standard gauge transformations, and one must perform appropriate field redefinitions to put them in a canonical form~\cite{Ghoshal:1992:GaugeGeneralCoordinate,David:2000:U1GaugeInvariance, Coletti:2003:AbelianNonabelianVector, Asano:2007:NewCovariantGauges, Asano:2012:ClosedStringField}.
One possible strategy is to start with the field whose gauge transformation has the lowest power of momentum, and then modify the other fields to make them invariant.

This section provides the new results of this paper: while the levels $(0, 0)$ and $(1, 0)$, and the odd sector of $(1, 1)$ have been worked out in~\cite{Okawa:2022:ClosedStringField}, we use a different approach to show that the new fields are not physical (in flat space).
Moreover, we also consider the even sector of the level $(1, 1)$, as well as the level $(2, 0)$.
CFT formulas used in this section are gathered in \Cref{app:formulas}.

We have made the computations with Cadabra~\cite{Peeters:2007:IntroducingCadabraSymbolic,Peeters:2018:Cadabra2ComputerAlgebra}.
The code can be found at \href{https://github.com/teaduality/closed-sft-without-level-matching}{\nolinkurl{github:teaduality/closed-sft-without-level-matching}}.

\subsection{Level expansion}

Let's start by focusing first on the SFT with level-matching (\Cref{sec:sft:level-matched}).
We decompose the string field $\Psi$ in eigenstates of the level operators $(N, \bar N)$ and of the momentum operator $p$ as:
\begin{equation}
	\ket{\Psi}
		= \int \frac{\dd^d k}{(2\pi)^d} \, \ket{\Psi(k)},
	\qquad
	\ket{\Psi(k)}
		= \sum_{\ell, \bar\ell \ge 0} \ket{\Psi_{\ell, \bar\ell}(k)},
\end{equation}
where
\begin{equation}
	\begin{gathered}
		p \ket{\Psi_{\ell, \bar\ell}(k)}
			= k \ket{\Psi_{\ell, \bar\ell}(k)},
		\\
		N \ket{\Psi_{\ell, \bar\ell}(k)}
			= \ell \ket{\Psi_{\ell, \bar\ell}(k)},
		\qquad
		\bar N \ket{\Psi_{\ell, \bar\ell}(k)}
			= \bar\ell \ket{\Psi_{\ell, \bar\ell}(k)}.
	\end{gathered}
\end{equation}
This implies:
\begin{equation}
	L_0^+ \ket{\Psi_{\ell, \bar\ell}(k)}
		= \left( \frac{\alpha' k^2}{2} \, + \ell + \bar\ell - 2 \right),
	\qquad
	L_0^- \ket{\Psi_{\ell, \bar\ell}(k)}
		= (\ell - \bar\ell) \ket{\Psi_{\ell, \bar\ell}(k)}.
\end{equation}
We will often omit the momentum dependence when there is no ambiguity.
The BPZ conjugate are defined as:
\begin{equation}
	\bra{\Psi_{\ell, \bar\ell}(k)}
		:= \bpz{\ket{\Psi_{\ell, \bar\ell}(k)}}.
\end{equation}
The reality condition \eqref{eq:sft:reality} becomes
\begin{equation}
    \adj{\ket{\Psi_{\ell, \bar\ell}(k)}}
        = - \bra{\Psi_{\ell, \bar\ell}(-k)}.
\end{equation}

Since $N$ and $\bar N$ commute with the BRST charge and $c_0^-$, the fields at different levels are orthogonal, such that the action splits into a single sum over $\ell$ and $\bar\ell$ of $S[\Psi_{\ell, \bar\ell}]$.
Moreover, fields with different worldsheet parity are orthogonal, so it is possible to separate each term $S_{\ell, \bar\ell}$ even further into even and odd sectors.
The final simplification is achieved by noting that the inner-product is proportional to $\delta^{(d)}(k + k')$ from \eqref{eq:normalization-bpz}, such that there is a single integral over momentum.
Thus, the action can be written as:
\begin{equation}
	\label{eq:exp:full-action}
    \begin{gathered}
	S
		= \sum_{\ell, \bar\ell \ge 0} \Big(
            S_{\ell, \bar\ell}^+
            + S_{\ell, \bar\ell}^-
            \Big),
	\\
    S_{\ell, \bar\ell}^\pm
		:= \frac{1}{2 V}
			\int \frac{\dd^d k}{(2\pi)^d} \,
			\bra{\Psi_{\ell, \bar\ell}^\pm(-k)} c_0^- Q_B \ket{\Psi_{\ell, \bar\ell}^\pm(k)},
    \end{gathered}
\end{equation}
where $V$ is the spacetime volume (to cancel $\delta^{(d)}(0)$ from the momentum inner-product), and $\Psi^+$ and $\Psi^-$ are respectively even and odd under worldsheet parity:
\begin{equation}
    \Omega \ket{\Psi_{\ell, \bar\ell}^\pm(k)}
        = \pm \ket{\Psi_{\ell, \bar\ell}^\pm(k)}.
\end{equation}
We will omit the parity index for levels where the field has a definite parity, and in generic formulas below.

The equation of motion of a given component is simply:
\begin{equation}
    Q_B \ket{\Psi_{\ell, \bar\ell}}
        = 0.
\end{equation}
The gauge parameters are expanded in the same manner:
\begin{equation}
	\ket{\Lambda}
		= \int \frac{\dd^d k}{(2\pi)^d} \, \ket{\Lambda(k)},
	\qquad
	\ket{\Lambda(k)}
		= \sum_{\ell, \bar\ell \ge 0} \ket{\Lambda_{\ell, \bar\ell}(k)},
\end{equation}
and also satisfy the reality condition:
\begin{equation}
    \adj{\ket{\Lambda_{\ell, \bar\ell}(k)}}
        = \bra{\Lambda_{\ell, \bar\ell}(-k)}.
\end{equation}
Then, string field components transform as:
\begin{equation}
    \delta_\Lambda \ket{\Psi_{\ell, \bar\ell}}
        = Q_B \ket{\Lambda_{\ell, \bar\ell}}.
\end{equation}
The same considerations apply for the parameter $\Omega$ in \eqref{eq:sft:gauge-transf-Lambda}, but it will be non-trivial only for the even sector of the level $(1, 1)$ (for other levels, it is not possible to satisfy all the constraints).

In the theory without level-matching (\Cref{sec:sft:unconstrained}), we also expand $\widetilde{\Psi}$ and $\widetilde{\Lambda}$ in level, momentum, and parity eigenstates.
All formulas are similar to the one above, except for the action~\cite{Okawa:2022:ClosedStringField}:
\begin{subequations}
\begin{gather}
	S_{\text{ext}}
		= \sum_{\ell, \bar\ell \ge 0} \Big(
            S_{\text{ext}, \ell, \bar\ell}^+
            + S_{\text{ext}, \ell, \bar\ell}^-
            \Big),
	\\
    S_{\ell, \bar\ell}^\pm
		:= \frac{1}{V}
			\int \frac{\dd^d k}{(2\pi)^d} \, \left[
                \frac{1}{2} \, \bra{\widetilde{\Psi}_{\ell, \bar\ell}^\mp(-k)} Q_B B \ket{\widetilde{\Psi}_{\ell, \bar\ell}^\mp(-k)}
                + \bra{\widetilde{\Psi}_{\ell, \bar\ell}^\mp(-k)} Q_B \ket{\Psi_{\ell, \bar\ell}^\pm(-k)}
                \right],
\end{gather}
\end{subequations}
and reality condition:
\begin{equation}
    \adj{\ket{\Psi_{\ell, \bar\ell}(k)}}
        = - \bra{\Psi_{\ell, \bar\ell}(-k)},
	\qquad
	\adj{\ket{\widetilde{\Psi}_{\ell, \bar\ell}(k)}}
		= \bra{\widetilde{\Psi}_{\ell, \bar\ell}(-k)}.
\end{equation}
The even-parity component of $\widetilde{\Psi}$ couples with the odd-parity part of $\Psi$ because of the difference in ghost numbers.
By convention, the sign superscript on the action corresponds to the parity of the physical string field $\Psi$.
Note that the first vanishes for fields which are not level-matched since $B$ contains $\delta(L_0^-)$.

We will not discuss the equations of motion for the spurious field $\widetilde{\Psi}$, with the exception of the level $(0, 0)$ and $(1, 0)$ as illustrations of the field redefinitions (see~\cite{Okawa:2022:ClosedStringField} for a detailed analysis).
For the same reason, we do not search for field redefinitions of those fields.

\subsection{Level (0, 0) -- level-matched tachyonic fields}
\label{sec:level-exp:00}

\subsubsection{Fields}

\paragraph{Physical string field}

The string field reads:
\begin{equation}
	\ket{\Psi_{0,0}(k)}
		= T(k) \ket{k, \downarrow\downarrow},
\end{equation}
where $T(k)$ is the tachyon, and $\ket{k, \downarrow\downarrow} := c_1 \bar c_1 \ket{k, 0}$ is the ghost energy vacuum, built on top of the $\mathrm{SL}(2, \C)$ vacuum (\Cref{app:formulas}).
The Euclidean and BPZ conjugates are:
\begin{equation}
	\adj{\ket{\Psi_{0,0}(k)}}
		= - \conj{T(k)} \bra{k, -},
	\qquad
	\bra{\Psi_{0,0}(k)}
		= T(k) \bra{- k, -},
\end{equation}
such that the reality condition \eqref{eq:sft:reality-ext} implies:
\begin{equation}
	\conj{T(k)}
		= T(-k).
\end{equation}

\paragraph{Spurious string field}

The spurious field is:
\begin{equation}
    \ket{\widetilde{\Psi}_{0,0}}
		= \Big(
			A_+(k) \, c_0^+
			+ A_-(k) \, c_0^-
			\Big) \ket{k, \downarrow\downarrow}.
\end{equation}
The Euclidean and BPZ conjugates are:
\begin{equation}
	\begin{aligned}
	\adj{\ket{\widetilde{\Psi}_{0,0}(k)}}
		&
		= - \bra{k, -} \Big(
			\conj{A_+(k)} \, c_0^+
			+ \conj{A_-(k)} \, c_0^-
			\Big),
	\\
	\bra{\widetilde{\Psi}_{0,0}(k)}
		&
		= - \bra{- k, -} \Big(
			A_+(k) \, c_0^+
			+ A_-(k) \, c_0^-
			\Big),
	\end{aligned}
\end{equation}
and the reality condition \eqref{eq:sft:reality-ext} implies that the spacetime fields are real:
\begin{equation}
	\conj{T(k)}
		= T(-k).
\end{equation}
For the other levels, we will not work out the BPZ and Euclidean conjugates nor the reality condition for the physical and spurious fields, since it is simple to compute them.

\subsubsection{Gauge invariance and field redefinition}

\paragraph{Physical string field}

The string field has no gauge transformation:
\begin{equation}
    \delta \ket{\Psi_{0,0}(k)}
        = 0
    \quad \Longrightarrow \quad
    \delta T(k)
        = 0.
\end{equation}

\paragraph{Spurious string field}

The spurious field has the following gauge parameter:
\begin{equation}
    \ket{\widetilde{\Lambda}_{(0,0)}(k)}
		= \alpha(k) \ket{k,-},
\end{equation}
with $\alpha(k)$ real.
We get the gauge transformations by acting with $Q_B$:
\begin{equation}
    \delta \ket{\widetilde{\Psi}_{(0,0)}}
		= Q_B \ket{\widetilde{\Lambda}_{(0,0)}}
		= \left(\frac{\alpha' k^2}{2} \, - 2 \right) \alpha(k) \, c_0^+ \ket{k, \downarrow\downarrow},
\end{equation}
which gives:
\begin{equation}
	\label{eq:exp:00-gauge-inv-aux}
    \delta A_+(k)
		= \left(\frac{\alpha' k^2}{2} \, - 2 \right) \alpha(k),
	\qquad
	\delta A_{-}(k)
		= 0.
\end{equation}

\subsubsection{Equation of motion}

\paragraph{Physical string field}

We have
\begin{equation}
	Q_B \ket{\Psi_{0,0}}
		= \left( \frac{\alpha' k^2}{2} \, - 2 \right) T(k) \, c_0^+ \ket{k, \downarrow\downarrow},
\end{equation}
which gives:
\begin{equation}
	\label{eq:exp:00-eom-phys}
	0
		= \left( k^2 - \frac{4}{\alpha'} \right) T(k).
\end{equation}
One recognizes the equation of motion for a tachyonic scalar with mass $m^2 = - 4 / \alpha'$.

\paragraph{Spurious string field}

We have
\begin{equation}
    Q_B \ket{\widetilde{\Psi}_{0,0}}
		= \left(- \frac{\alpha' k^2}{2} \, + 2 \right) A_-(k) \, c_0^-c_0^+ \ket{k,-},
\end{equation}
which gives
\begin{equation}
	0
    	= \left(k^2 - \frac{4}{\alpha'} \right) A_-(k).
\end{equation}
This is also the equation of motion of a tachyonic scalar of the same mass.
The field $A_+$ has no equation of motion, and it is pure gauge off-shell, according to \eqref{eq:exp:00-gauge-inv-aux}.

\subsubsection{Action}
\label{sec:level-exp:00:action}

The action reads:
\begin{equation}
    S_{0,0}
		= - \int \frac{\dd^d k}{(2\pi)^d} \left[
			\frac{1}{2} \, A_-(-k) \left(\frac{\alpha' k^2}{4} - 1 \right) A_-(k)
			+ A_-(-k) \left(\frac{\alpha' k^2}{4} - 1\right) T(k)
			\right].
\end{equation}
We see that the auxiliary field $A_+$ does not appear inside, which is expected since it does not have any equation of motion.
The reason is that $T$ and  $A_+$ are odd under the worldsheet parity, and $A_-$ is even: from \eqref{eq:exp:full-action}, it implies that $A_-$ cannot couple to any physical field.

Setting $A_- = - T$ as given in \eqref{eq:sft:field-aux-to-phys} gives the usual action for the tachyon in Euclidean signature:
\begin{equation}
	S_{0,0}
		= \frac{\alpha'}{8}
			\int \frac{\dd^d k}{(2\pi)^d} \,
			T(-k) \left( k^2 - \frac{4}{\alpha'} \right) T(k).
\end{equation}
Varying this equation gives the equation \eqref{eq:exp:00-eom-phys} as needed.
We see that the action can be canonically normalized if one defines $T \to 2 T / \sqrt{\alpha'}$.
Note that the action above is not obtained by integrating out $A_- = - 2 T$, since it gives a vanishing action.

\subsection{Level (1, 0) -- non-level-matched tachyonic fields}
\label{sec:level-exp:10}

\subsubsection{Fields}

\paragraph{Physical string field}

The string field reads:\footnotemark{}
\footnotetext{%
	We use the basis $(c_0^+, c_0^-)$, while~\cite{Okawa:2022:ClosedStringField} uses $(c_0, \bar c_0)$.
}%
\begin{equation}
	\ket{\Psi_{1,0}(k)}
		= \Big(
			\I D_\mu(k) \, \alpha_{-1}^\mu
			+ C_+(k) \, c_0^+ b_{-1}
			+ C_-(k) \, c_0^- b_{-1}
			\Big) \ket{k, \downarrow\downarrow}.
\end{equation}
All fields are correctly real.

\paragraph{Spurious string field}

The spurious field is:
\begin{equation}
     \ket{\widetilde{\Psi}_{1,0}}
	 	= \Big(
			E(k) \, c_0^- c_0^+ b_{-1}
			+ J(k) \, c_{-1}
			+ \I F_{\mu}(k) \, c_0^- \alpha_{-1}^{\mu}
			+ \I H_{\mu}(k) \, c_0^+ \alpha_{-1}^{\mu}
			\Big) \ket{k, \downarrow\downarrow}.
\end{equation}

\subsubsection{Gauge invariance and field redefinition}

\paragraph{Physical string field}

The gauge parameter reads:
\begin{equation}
	\ket{\Lambda_{1,0}(k)}
		= \lambda(k) \, b_{-1} \ket{k, \downarrow\downarrow}.
\end{equation}
The action of $Q_B$ gives
\begin{equation}
	Q_B \ket{\Lambda_{1,0}}
		= \lambda(k) \, \bigg[
			\bigg( \frac{\alpha'}{2} \, k^2 - 1 \bigg) c_0^+ b_{-1}
			+ c_0^- b_{-1}
			+ \sqrt{\frac{\alpha'}{2}} \, k \cdot \alpha_{-1}
			\bigg] \ket{k, \downarrow\downarrow},
\end{equation}
which gives the gauge transformations of each spacetime fields:
\begin{equation}
	\begin{gathered}
	\delta D_\mu(k)
		= - \sqrt{\frac{\alpha'}{2}} \, \I k_\mu \lambda(k),
	\qquad
	\delta C_+(k)
		= \bigg( \frac{\alpha'}{2} \, k^2 - 1 \bigg) \lambda(k),
	\qquad
	\delta C_-(k)
		= \lambda(k).
	\end{gathered}
\end{equation}

In the gauge transformations above, $C_-$ has no power of $k$, which suggests the following field redefinitions:
\begin{equation}
	\label{eq:exp:01-field-redef}
	\begin{gathered}
	\bar C_-(k)
		= C_-(k),
	\qquad
	\bar C_+(k)
		= C_+(k)
			+ C_-(k)
			- \sqrt{\frac{\alpha'}{2}} \, \I k^\mu D_\mu(k),
	\\\
	\bar D_\mu(k)
		= D_\mu(k)
			+ \sqrt{\frac{\alpha'}{2}} \, \I k_\mu C_-(k),
	\end{gathered}
\end{equation}
such that
\begin{equation}
	\delta \bar C_-(k)
		= \lambda(k),
	\qquad
	\delta \bar C_+(k)
		= \delta \bar D_\mu(k)
		= 0.
\end{equation}
Note that $C_+$ is a Nakanishi--Lautrup (NL) field for $D_\mu$, but it is not interesting to exploit this property here.
In terms of these new variables, the string field is:
\begin{equation}
	\begin{aligned}
	\ket{\Psi_{1,0}(k)}
		&
		= \bigg[
			\I \bar D_\mu(k) \bigg(
				\alpha_{-1}^\mu
				+ \sqrt{\frac{\alpha'}{2}} \, k^\mu \, c_0^+ b_{-1}
				\bigg)
			+ \bar C_+(k) \, c_0^+ b_{-1}
			\\ & \qquad
			+ \bar C_-(k) \bigg(
				\sqrt{\frac{\alpha'}{2}} \, k_\mu \alpha_{-1}^\mu
				+ \bigg( \frac{\alpha'}{2} \, k^2 - 1 \bigg) c_0^+ b_{-1}
				+ c_0^- b_{-1}
				\bigg)
			\bigg] \ket{k, \downarrow\downarrow}.
	\end{aligned}
\end{equation}

\paragraph{Spurious string field}

The gauge parameter of the spurious field is:
\begin{equation}
    \ket{\widetilde{\Lambda}_{1,0}(k)}
		= \Big(
			\gamma(k) \, c_0^+ b_{-1}
			+ \zeta(k) \, c_0^- b_{-1}
			+ \I \epsilon_{\mu}(k) \, \alpha^{\mu}_{-1}
			\Big) \ket{k, \downarrow\downarrow}.
\end{equation}

The gauge parameter for the gauge transformation is
\begin{equation}
	\ket{\widetilde \Omega_{1,0}(k)}
		= \sigma(k) \, b_{-1} \ket{k, \downarrow\downarrow},
\end{equation}
such that
\begin{equation}
	\delta \gamma(k)
		= \left( \frac{\alpha' k^2}{2} - 1 \right) \sigma(k),
	\qquad
	\delta \zeta(k)
		= \sigma(k),
	\qquad
	\delta \epsilon_{\mu}(k)
		= - \sqrt{\frac{\alpha'}{2}} \, \I k_{\mu} \sigma(k).
\end{equation}
This suggests the following redefinitions for the parameters:
\begin{equation}
	\begin{gathered}
	\bar \gamma(k)
		= \gamma(k)
			- \left( \frac{\alpha' k^2}{2} - 1 \right) \zeta(k),
	\qquad
	\bar \zeta(k)
		= \zeta(k),
	\\
	\bar \epsilon_{\mu}(k)
		= \epsilon_{\mu}(k)
			+ \sqrt{\frac{\alpha'}{2}} \, \I k_{\mu} \zeta(k).
	\end{gathered}
\end{equation}

The action of $Q_B$ gives the following gauge transformations:
\begin{equation}
	\begin{aligned}
    \delta E(k)
		&
		= \gamma(k)
			- \left(\frac{\alpha' k^2}{2} - 1 \right) \zeta(k),
	\\
    \delta J(k)
		&
		= - \gamma(k)
			- \zeta(k)
			+ \sqrt{\frac{\alpha'}{2}} \, \I k^{\mu} \epsilon_{\mu}(k),
	\\
    \delta F_{\mu}(k)
		&
		= \epsilon_{\mu}(k)
			+ \sqrt{\frac{\alpha'}{2}} \, \I k_{\mu} \zeta(k),
	\\
    \delta H_{\mu}(k)
		&
		= \left( \frac{\alpha' k^2}{2} - 1 \right) \epsilon_{\mu}(k)
			+ \sqrt{\frac{\alpha'}{2}} \, \I k_{\mu} \gamma(k).
	\end{aligned}
\end{equation}
We perform the change of variables for the gauge parameters, gauge fixing $\zeta = 0$, which gives:
\begin{equation}
	\begin{gathered}
    \delta E(k)
		= \bar \gamma(k),
	\qquad
    \delta J(k)
		= - \bar \gamma(k)
			+ \sqrt{\frac{\alpha'}{2}} \, \I k^{\mu} \bar \epsilon_{\mu}(k),
	\\
    \delta F_{\mu}(k)
		= \bar \epsilon_{\mu}(k),
	\qquad
    \delta H_{\mu}(k)
		= \left( \frac{\alpha' k^2}{2} - 1 \right) \bar \epsilon_{\mu}(k)
			+ \sqrt{\frac{\alpha'}{2}} \, \I k_{\mu} \bar \gamma(k).
	\end{gathered}
\end{equation}

We perform the field redefinitions
\begin{equation}
	\label{eq:exp:01-field-aux-redef}
	\begin{gathered}
	\bar E(k)
		= E(k),
	\qquad
	\bar J(k)
		= E(k)
			+ J(k)
			- \sqrt{\frac{\alpha'}{2}} \, \I k^{\mu} F_{\mu}(k),
	\\
	\bar F_{\mu}(k)
		= F_{\mu}(k)
	\qquad
	\bar H_{\mu}(k)
		= H_{\mu}(k)
			- \left( \frac{\alpha' k^2}{2} - 1 \right) F_{\mu}(k)
			- \sqrt{\frac{\alpha'}{2}} \, \I k_{\mu} E(k),
	\end{gathered}
\end{equation}
such that
\begin{equation}
	\begin{gathered}
    \delta \bar E(k)
		= \bar \gamma(k),
	\qquad
	\delta \bar F_{\mu}(k)
		= \bar \epsilon_{\mu}(k),
	\qquad
    \delta \bar J(k)
    	= \delta \bar H_{\mu}(k)
		= 0.
	\end{gathered}
\end{equation}

\subsubsection{Equation of motion}

\paragraph{Physical string field}

We have
\begin{equation}
\begin{aligned}
	Q_B \ket{\Psi_{1,0}}
		&
		=
			\bigg[
				\bigg( \frac{\alpha'}{2} \, k^2 - 1 \bigg) \I D_\mu(k)
				- \sqrt{\frac{\alpha'}{2}} \, k_\mu C_+(k)
				\bigg] \alpha_{-1}^\mu c_0^+ \ket{k, \downarrow\downarrow}
			\\ & \qquad
			+
			\bigg[
				\I D_\mu(k)
				- \sqrt{\frac{\alpha'}{2}} \, k_\mu C_-(k)
				\bigg] \alpha_{-1}^\mu c_0^- \ket{k, \downarrow\downarrow}
			\\ & \qquad
			+
			\bigg[
				C_+(k)
				- \bigg( \frac{\alpha'}{2} \, k^2 - 1 \bigg) C_-(k)
				\bigg] c_0^- c_0^+ b_{-1} \ket{k, \downarrow\downarrow}
			\\ & \qquad
			+
			\bigg[
				\sqrt{\frac{\alpha'}{2}} \, \I k^\mu D_\mu(k)
				- C_+(k)
				- C_-(k)
				\bigg] c_{-1} \ket{k, \downarrow\downarrow},
	\end{aligned}
\end{equation}
which gives the equations of motion:
\begin{equation}
	\label{eq:exp:01-eom-phys}
	\begin{gathered}
		0
			= \bigg( \frac{\alpha'}{2} \, k^2 - 1 \bigg) D_\mu(k)
				+ \sqrt{\frac{\alpha'}{2}} \, \I k_\mu C_+(k),
		\qquad
		0
			= D_\mu(k)
				+ \sqrt{\frac{\alpha'}{2}} \, \I k_\mu C_-(k),
		\\
		0
			= \bigg( \frac{\alpha'}{2} \, k^2 - 1 \bigg) C_-(k)
				- C_+(k),
		\qquad
		0
			= \sqrt{\frac{\alpha'}{2}} \, \I k^\mu D_\mu(k)
				- C_+(k)
				- C_-(k).
	\end{gathered}
\end{equation}

In terms of the redefined fields \eqref{eq:exp:01-field-redef}, the equations of motion become, after simplification:
\begin{equation}
	0
		= \bar C_+(k)
		= \bar D_\mu(k).
\end{equation}
Note that $\bar C_-(k)$ does not appear anymore, which is normal since it can be completely gauge fixed, this shows that there are no dynamical degrees of freedom as expected.

\paragraph{Spurious string field}

The equations of motion are:
\begin{equation}
	\begin{aligned}
		0
			&
			= J(k)
				+ E(k)
				+ \sqrt{\frac{\alpha'}{2}} \, \I k^{\mu} F_{\mu}(k),
		\\
		0
			&
			= H_{\mu}(k)
				- \left(\frac{\alpha'k^2}{2} - 1 \right) F_{\mu}(k)
				- \sqrt{\frac{\alpha'}{2}} \, \I k_{\mu} E(k),
		\\
		0
			&
			= - \left(\frac{\alpha'k^2}{2} - 1 \right) J(k)
				+ E(k)
				+ \sqrt{\frac{\alpha'}{2}} \, \I k^{\mu} H_{\mu}(k).
	\end{aligned}
\end{equation}
We can now perform the field redefinitions and set $\bar E = \bar H_\mu = 0$ with a gauge transformation:
\begin{equation}
	\begin{gathered}
		0
			= \left(\frac{\alpha'k^2}{2} - 1 \right) J(k),
		\qquad
		0
			= \left(\frac{\alpha'k^2}{2} - 1 \right) F_{\mu}(k),
		\\
		0
			= J(k)
				+ \sqrt{\frac{\alpha'}{2}} \, \I k^{\mu} F_{\mu}(k).
	\end{gathered}
\end{equation}
This corresponds to free equations of motion for $J$ and $F_\mu$ as it should for the spurious fields.
The additional constraint means that $J$ describes the spin $0$ component of the massive vector field $F_\mu$.

\subsubsection{Action}

The action is
\begin{equation}
	\begin{aligned}
    S
		= \frac{1}{2} \int\frac{\dd^d k}{(2\pi)^d} \bigg[
			&
			E(-k) \bigg(
				- C_+(k)
				- C_-(k)
				+ \sqrt{\frac{\alpha'}{2}} \, \I k^{\mu} D_{\mu}(k)
				\bigg)
			\\ &
			+ J(-k) \bigg(
				 - C_+(k)
				 + \Big( \frac{\alpha' k^2}{2} - 1 \Big) C_-(k)
				 \bigg)
			\\ &
			- F^{\mu}(-k) \bigg(
				\Big( \frac{\alpha' k^2}{2} - 1 \Big) D_{\mu}(k)
				+ \sqrt{\frac{\alpha'}{2}} \, \I k_{\mu} C_+(k)
				\bigg)
			\\ &
			+ H^{\mu}(-k) \bigg(
				D_{\mu}(k)
				+ \sqrt{\frac{\alpha'}{2}} \, \I k_{\mu} C_-(k)
				\bigg)
			\bigg].
	\end{aligned}
\end{equation}
Note that we cannot integrate out the spurious field to get an action reproducing the equations of motion \eqref{eq:exp:01-eom-phys}.
Performing the field redefinitions \eqref{eq:exp:01-field-redef} and \eqref{eq:exp:01-field-aux-redef}, the action becomes:
\begin{equation}
	\begin{aligned}
    S
		= \frac{1}{2} \int\frac{\dd^d k}{(2\pi)^d} \bigg[
			&
			- \bar E(-k) \bar C_+(k)
			- \sqrt{\frac{\alpha'}{2}} \, \bar F_{\mu}(-k) \I k_{\mu} \bar C_+(k)
			\\
			&
			- \bigg(
					\bar J(-k)
					- \bar E(-k)
					- \sqrt{\frac{\alpha'}{2}} \, \I k^{\mu} \bar F_{\mu}(-k)
				\bigg)
				\\ & \hspace{2cm}
				\times
				\bigg(
					\bar C_+(k)
					+ \sqrt{\frac{\alpha'}{2}} \, \I k^\mu \bar D_\mu(k)
					\bigg)
			\\
			&
			+ \bigg(
				\bar H_{\mu}(-k)
				+ \Big( \frac{\alpha' k^2}{2} - 1 \Big) \bar F_{\mu}(-k)
				- \sqrt{\frac{\alpha'}{2}} \, \I k_{\mu} \bar E(-k)
				\bigg) \bar D_\mu(k)
			\bigg].
	\end{aligned}
\end{equation}
Note that $\bar C_-$ does not appear at all in the action.

\subsection{Level (1, 1) odd -- level-matched massless fields}
\label{sec:level-exp:11-odd}

\subsubsection{Fields}

\paragraph{Physical string field}

The string field reads:
\begin{equation}
	\begin{aligned}
	\ket{\Psi_{1,1}^-(k)}
		&
		= \Big(
			G_{\mu\nu}(k) \, (\alpha_{-1}^\mu \bar \alpha_{-1}^\nu+\alpha_{-1}^\nu \bar \alpha_{-1}^\mu)
			+ \I E^-_\mu(k) \, c_0^+ \big( \alpha_{-1}^\mu \bar b_{-1} + b_{-1} \bar \alpha_{-1}^\mu \big)
			\\ & \qquad
			+ \I A^-_\mu(k) \, c_0^- \big( \alpha_{-1}^\mu \bar b_{-1} - b_{-1} \bar \alpha_{-1}^\mu \big)
			+ D(k) \, \big( b_{-1} \bar c_{-1} - c_{-1} \bar b_{-1} \big)
			\\ & \qquad
			+ B(k) \, c_0^- c_0^+ b_{-1} \bar b_{-1}
			\Big) \ket{k, \downarrow\downarrow},
	\end{aligned}
\end{equation}
where $G_{\mu\nu} = G_{\nu\mu}$ corresponds to the metric, $D$ to the ghost-dilaton, and $E^-_\mu$ to the NL field associated with diffeomorphisms.
All fields are correctly real.

\paragraph{Spurious string field}

The spurious field is:
\begin{equation}
	\begin{aligned}
	\ket{\widetilde{\Psi}_{1,1}^+}
		= \Big(&
			M(k) \, (b_{-1} \bar{c}_{-1} - c_{-1} \bar{b}_{-1}) c_0^-
			+ N(k) \, (b_{-1} \bar{c}_{-1} + c_{-1} \bar{b}_{-1}) c_0^+
			\\ &
			+ \I P_{\mu}(k) \, (\alpha^{\mu}_{-1} \bar{b}_{-1} + b_{-1} \bar{\alpha}^{\mu}_{-1})c_0^-c_0^+
			+ \I Q_{\mu}(k) \, (\alpha^{\mu}_{-1} \bar{c}_{-1} - c_{-1} \bar{\alpha}^{\mu}_{-1})
			\\ &
			+ R_{\mu\nu}(k) \, (\alpha^{\mu}_{-1} \bar{\alpha}^{\nu}_{-1} + \alpha^{\nu}_{-1} \bar{\alpha}^{\mu}_{-1}) c_0^-
			\\ &
			+ S_{\mu\nu}(k) \, (\alpha^{\mu}_{-1} \bar{\alpha}^{\nu}_{-1} - \alpha^{\nu}_{-1} \bar{\alpha}^{\mu}_{-1}) c_0^+
			\Big) \ket{k, \downarrow \downarrow}.
	\end{aligned}
\end{equation}
All fields are correctly real.

\subsubsection{Gauge invariance and field redefinition}

\paragraph{Physical string field}

The gauge parameter reads:
\begin{equation}
	\begin{aligned}
	\ket{\Lambda_{1,1}^-(k)}
		&
		= \Big(
			\chi^-(k) \, c_0^- b_{-1} \bar b_{-1}
			+ \I \xi^-_\mu(k) \big( \alpha_{-1}^\mu \bar b_{-1} + b_{-1} \bar \alpha_{-1}^\mu \big)
			\Big) \ket{k, \downarrow\downarrow},
	\end{aligned}
\end{equation}
and its components are real.

The gauge transformations we obtain by acting with $Q_B$ are
\begin{equation}
	\label{eq:exp:11o-gauge-inv-aux}
	\begin{gathered}
		\delta B(k)
			= - \frac{\alpha' k^2}{2} \, \chi^-(k),
		\qquad
		\delta E^-_{\mu}(k)
			= - \frac{\alpha' k^2}{2} \, \xi^-_{\mu}(k),
		\\
		\delta D(k)
			= \chi^-(k) - \sqrt{\frac{\alpha'}{2}} \, \I k^{\mu} \xi^-_{\mu}(k),
		\qquad
		\delta A^-_{\mu}(k)
			= - \sqrt{\frac{\alpha'}{2}} \, \I k_{\mu} \chi^-(k),
		\\
		\delta G_{\mu\nu}(k)
			= \frac{1}{2} \, \sqrt{\frac{\alpha'}{2}} \,
				\big( \I k_{\mu} \xi^-_{\nu}(k) + \I k_{\nu} \xi^-_{\mu}(k) \big).
	\end{gathered}
\end{equation}

With the same logic as in the $(1, 0)$ case, we perform the following field redefinition:
\begin{equation}
	\label{eq:exp:11o-field-aux-redef-B}
    \bar B(k)
		= B(k) + \sqrt{\frac{\alpha'}{2}} \, \I k^{\mu} A^-_{\mu}(k),
\end{equation}
which renders $\bar B(k)$ gauge invariant.
Next, we also redefine the ghost-dilaton and the NL fields as~\cite{Yang:2005:ClosedStringTachyon,Asano:2012:ClosedStringField}
\begin{equation}
    \begin{gathered}
	\bar D(k)
		= D(k) + G_{\mu}^{\mu}(k),
	\\
	\bar E^-_{\mu}(k)
		= E^-_{\mu}(k)
			- A^-_{\mu}(k)
			- 2 \sqrt{\frac{\alpha'}{2}} \, \I k^\nu G_{\mu\nu}(k)
			- \sqrt{\frac{\alpha'}{2}} \, \I k_\mu D(k),
	\end{gathered}
\end{equation}
such that $\bar D$ is invariant under reparametrizations generated by $\xi_\mu^-$, and $\bar E^-_\mu$ is completely gauge invariant.
With this parametrization, the metric is in the string frame~\cite{Yang:2005:ClosedStringTachyon}.
We could perform further field redefinitions to get the metric in the Einstein frame~\cite{Yang:2005:ClosedStringTachyon,Asano:2012:ClosedStringField}, but it is sufficient to keep the string frame.
Note that $\bar D^-$ still transforms under the transformation with parameter $\chi^-$, and in a way which says that it is pure gauge.
This is expected since the ghost-dilaton is non-trivial only in the semi-relative cohomology where $b_0^- = 0$~\cite{Bergman:1995:DilatonTheoremClosed,Belopolsky:1996:WhoChangesString,Astashkevich:1997:StringCenterMass}.
We could make $A^-_\mu$ gauge invariant and gauge fix $\chi^-$ by removing $D$, but it is simpler to keep the ghost-dilaton as a state and gauge fix $\chi^-$ with a condition on $A^-_\mu$.\footnotemark{}
\footnotetext{%
	\label{ft:dilaton}
	However, the zero-momentum mode $\chi^-(0)$ remains and can be used to gauge-fix the zero-momentum dilaton.
	As a result, it is not clear which operator changes the string coupling.
	We thank Ted Erler for discussions on this point.
}%
Hence, we do not change the remaining fields:
\begin{equation}
	\bar G_{\mu\nu}(k)
		= G_{\mu\nu}(k),
	\qquad
	\bar A^-_{\mu}(k)
		= A^-_{\mu}(k).
\end{equation}

\paragraph{Spurious string field}

\begin{equation}
\begin{aligned}
	\ket{\widetilde{\Lambda}_{1,1}^+(k)}
		= \Big(
			&
			\alpha(k) \, (b_{-1} \bar{c}_{-1} + c_{-1} \bar{b}_{-1})
			+ \I \beta_{\mu}(k) \, (\alpha^{\mu}_{-1} \bar{b}_{-1} - b_{-1} \bar{\alpha}^{\mu}_{-1}) \, c_0^+
			\\ &
			+ \I \gamma_{\mu}(k) \, (\alpha^{\mu}_{-1} \bar{b}_{-1} + b_{-1} \bar{\alpha}^{\mu}_{-1}) \, c_0^-
			\\ &
			+ \delta_{\mu\nu}(k) \, (\alpha_{-1}^{\mu} \bar{\alpha}_{-1}^{\nu} -\alpha^{\nu}_{-1} \bar{\alpha}_{-1}^{\mu})
			\Big) \ket{k,\downarrow \downarrow}.
\end{aligned}
\end{equation}
The gauge transformations we obtain by acting with $Q_B$ are
\begin{align}
	\nonumber
	\delta M(k)
		&
		= - \sqrt{\frac{\alpha'}{2}} \, \I k^{\mu} \gamma_{\mu}(k),
	\\
	\delta N(k)
		&
		= \frac{\alpha' k^2}{2} \, \alpha(k)
			+ \sqrt{\frac{\alpha'}{2}} \, \I k^{\mu} \beta_{\mu}(k)
	\\
	\nonumber
	\delta P_{\mu}(k)
		&
		=  \frac{\alpha' k^2}{2} \, \gamma_{\mu}(k),
	\\
	\nonumber
	\delta Q_{\mu}(k)
		&
		= \beta_{\mu}(k)
			- \gamma_{\mu}(k)
			-\I \sqrt{\frac{\alpha'}{2}} \, k_{\mu} \alpha(k)
			- 2\I \sqrt{\frac{\alpha'}{2}} \, k^{\nu} \delta_{\mu\nu}(k),
	\\
	\nonumber
	\delta R_{\mu\nu}(k)
		&
		= \frac{1}{2} \sqrt{\frac{\alpha'}{2}} \,
			\big(\I k_{\nu} \gamma_{\mu}(k) + \I k_{\mu} \gamma_{\nu}(k) \big),
	\\
	\delta S_{\mu\nu}(k)
		&
		= \frac{1}{2} \left[
			\sqrt{\frac{\alpha'}{2}} \, \big( \I k_{\nu} \beta_{\mu}(k) - \I k_{\mu} \beta_{\nu}(k) \big)
			+ \alpha' k^2 \delta_{\mu\nu}(k)
			\right].
\end{align}

\subsubsection{Equation of motion}

\paragraph{Physical string field}
Omitting computational details, $Q_B\ket{\Psi^-_{1,1}} = 0$ gives:
\begin{equation}
\begin{aligned}
	0
		&
		= \frac{\alpha' k^2}{2} \, D(k)
			- \sqrt{\frac{\alpha'}{2}} \, \I k^{\mu} E^-_{\mu}(k)
			+ B(k),
	\\
	0
		&
		= \frac{\alpha' k^2}{2} \, A^-_{\mu}(k)
			- \sqrt{\frac{\alpha'}{2}} \, \I k_{\mu} B(k),
	\\
	0
		&
		= \alpha' k^2 \, G_{\mu\nu}(k)
			+ \sqrt{\frac{\alpha'}{2}} \, \big(\I k_{\nu} E^-_{\mu}(k) + \I k_{\mu} E^-_{\nu}(k) \big),
	\\
	0
		&
		= E^-_{\mu}(k)
			- A^-_{\mu}(k)
			- \sqrt{\frac{\alpha'}{2}} \, \I k_{\mu}D(k)
			- \sqrt{2\alpha'} \, \I k^{\nu} G_{\mu\nu}(k),
	\\
	0
		&
		= B(k)
			+ \sqrt{\frac{\alpha'}{2}} \, \I k^{\mu} A^-_{\mu}(k),
	\\
	0
		&
		= \sqrt{\frac{\alpha'}{2}}\, \big( \I k_{\nu} A^-_{\mu}(k)- \I k_{\mu} A^-_{\nu}(k) \big).
\end{aligned}
\end{equation}
As usual, some equations are redundant and can be obtained by combining others and their products with $k_\mu$.
The penultimate equation shows that $B$ is a NL field for $A_\mu^-$, consistently with the transformation \eqref{eq:exp:11o-gauge-inv-aux}, motivating further the choice for the redefinition \eqref{eq:exp:11o-field-aux-redef-B}.

We can combine the 3rd and 4th equations of motion to get:
\begin{equation}
\begin{aligned}
	0
		&
		= \alpha' \Big[
				k^2 G_{\mu\nu}
				+ k_{\mu} k_{\nu} G_{\rho}^{\rho}
				- k_{\nu} k^{\rho} G_{\mu\rho}
				- k_{\mu} k^{\rho} G_{\nu\rho}
				- \eta_{\mu\nu} k^2 G_{\rho}^{\rho}
				+ \eta_{\mu\nu} k^{\rho} k^{\sigma} G_{\rho\sigma}
				\Big]
			\\ & \quad
			+\alpha' \left(\frac{k^2}{2} \, \eta_{\mu\nu} - k_{\mu} k_{\nu} \right) \bar D
			+ \sqrt{\frac{\alpha'}{2}} \,
				\big(\I k_{\nu} A^-_{\mu} + \I k_{\mu} A^-_{\nu}\big)
			- \sqrt{\frac{\alpha'}{2}} \, \I k^{\rho} A^-_{\rho} \eta_{\mu\nu}.
\end{aligned}
\end{equation}
The first line corresponds to the linearized Einstein equation.

The last equation tells that the field strength of $A_\mu^-$ is zero:
\begin{equation}
	F_{\mu\nu}(A^-)
		= 0,
\end{equation}
which implies $A_\mu^- = 0$ on flat space (the RHS of this equation stays zero even after including interaction, such that this field is genuinely non-propagating~\cite{Okawa:2022:ClosedStringField}).
However, on a toroidal background it is possible to have non-trivial fluxes, implying that new solutions may be found without the level-matching condition~\cite{Okawa:2022:ClosedStringField}.
This is the observation which sparked our interest in studying the complete free action, before being able to investigate properties of the background such as $T$-duality.
On flat space, the penultimate equation implies $B = 0$, such that the other equations reduce to the usual equations of motion for the ghost-dilaton, metric and NL vector field.
Hence, closed string theories with and without level-matching are equivalent at the perturbative level, but may be inequivalent non-perturbatively.

After performing the field redefinition, we get:
\begin{equation}
\begin{aligned}
	0
		&
		= \bar E^-_{\mu}(k)
		= \bar B(k),
	\\
	0
		&
		= \alpha' k^2 \, \bar D(k)
			+ \alpha' (\eta^{\mu\nu} k^2 + k^{\mu} k^{\nu}) \, \bar G_{\mu\nu}(k)
			- 2 \sqrt{\frac{\alpha'}{2}} \, \I k^{\mu} \bar A^-_{\mu}(k),
	\\
	0
		&
		= \alpha' k^2 \, \bar G_{\mu\nu}(k)
			- \alpha' \, \big(k_{\nu} k^\rho \bar G_{\mu\rho}(k) + k_{\mu} k^\rho \bar G_{\nu\rho}(k) \big)
			- \alpha' \, k_{\mu} k_\nu \bar D(k)
			\\ & \qquad
			+ \sqrt{\frac{\alpha'}{2}} \, \big(\I k_{\nu} A^-_{\mu}(k) + \I k_{\mu} \bar A^-_{\nu}(k) \big),
	\\
	0
		&
		= \sqrt{\frac{\alpha'}{2}}\, \big( \I k_{\nu} \bar A^-_{\mu}(k)- \I k_{\mu} \bar A^-_{\nu}(k) \big).
\end{aligned}
\end{equation}

\subsubsection{Action}

The action is:
\begin{equation}
\begin{aligned}
	S_{1,1}^-
		&
		= \int \frac{\dd^d k}{(2\pi)^d} \Bigg[
			M(-k) \, \frac{\alpha' k^2}{4} \, M(k)
			- \sqrt{\frac{\alpha'}{2}} \, P_{\mu}(-k) \, \I k^{\mu} M(k)
			\\ & \hspace{2cm}
			- \frac{1}{2} \, P^{\mu}(-k) P_{\mu}(k)
			- R_{\mu\nu}(-k) \, \frac{\alpha' k^2}{2}  \, R^{\mu\nu}(k)
			\\ & \hspace{2cm}
			+ \sqrt{\frac{\alpha'}{2}} \, R_{\mu\nu}(-k) \big(
				\I k_{\nu} P_{\mu}(k)+ \I k_{\mu} P_{\nu}(k)
				\big)
			\Bigg]
		\\ & \quad
		+ \int \frac{\dd^d k}{(2\pi)^d} \Bigg[
			M(-k) \bigg(
				\frac{\alpha' k^2}{2} \, D(k)
				- \sqrt{\frac{\alpha'}{2}} \, \I k^{\mu} E^-_{\mu}(k)
				+ B(k)
				\bigg)
			\\ & \hspace{2.5cm}
			+ N(-k) \bigg(
				B(k)+\I\sqrt{\frac{\alpha'}{2}}k^{\mu}A^-_{\mu}(k)
				\bigg)
			\\ & \hspace{2.5cm}
			+ P^{\mu}(-k) \bigg(
				E^-_{\mu}(k)-A^-_{\mu}(k)-\I\sqrt{\frac{\alpha'}{2}}k_{\mu}D(k)-2\I G_{\mu\nu}(k)\sqrt{\frac{\alpha'}{2}}k^{\nu}
				\bigg)
			\\ & \hspace{2.5cm}
			- Q^{\mu}(-k) \bigg(
				\frac{\alpha' k^2}{2} \, A^-_{\mu}(k)
					- \sqrt{\frac{\alpha'}{2}} \, \I k_{\mu} B(k)
				\bigg)
			\\ & \hspace{2.5cm}
			- R^{\mu\nu}(-k) \bigg(
				\alpha' k^2 \, G_{\mu\nu}(k)
				+ \sqrt{\frac{\alpha'}{2}} \, \big(\I k_{\nu} E^-_{\mu}(k) + \I k_{\mu} E^-_{\nu}(k) \big)
				\bigg)
			\\ & \hspace{2.5cm}
			+ S^{\mu\nu}(-k) \bigg(
				\sqrt{\frac{\alpha'}{2}}\, \big( \I k_{\nu} A^-_{\mu}(k)- \I k_{\mu} A^-_{\nu}(k) \big)
				\bigg)
				\Bigg].
\end{aligned}
\end{equation}

\subsection{Level (1, 1) even -- level-matched massless fields}

\subsubsection{Fields}

\paragraph{Physical string field}

The string field reads:
\begin{equation}
	\begin{aligned}
    \ket{\Psi_{1,1}^+(k)}
		= \Big(
			& D^+(k) \, ( b_{-1}\bar{c}_{-1} + c_{-1}\bar{b}_{-1} )
			+ \I E^+_{\mu}(k) \, (\alpha^{\mu}_{-1} \bar{b}_{-1} - b_{-1}\bar{\alpha}^{\mu}_{-1} )c_0^+
			\\ &
			+ \I A^+_{\mu}(k) \, (\alpha^{\mu}_{-1} \bar{b}_{-1} + b_{-1} \bar{\alpha}^{\mu}_{-1}) c_0^-
			\\ &
			+ B_{\mu\nu}(k) \, (\alpha^{\mu}_{-1} \bar{\alpha}^{\nu}_{-1} - \alpha^{\nu}_{-1} \bar{\alpha}^{\mu}_{-1})
			\Big) \ket{k, \downarrow\downarrow}
\end{aligned}
\end{equation}
where $B_{\mu\nu} = -B_{\nu\mu}$ corresponds to the Kalb--Ramond field, and  $E^+_{\mu}$ its NL field.

\paragraph{Spurious string field}

The spurious field is:
\begin{equation}
	\begin{aligned}
	\ket{\widetilde{\Psi}_{1,1}^-}
		= \Big(
			&
			M'(k) \, (b_{-1} \bar{c}_{-1} + c_{-1} \bar{b}_{-1}) c_0^-
			+ N'(k) \, (b_{-1} \bar{c}_{-1} - c_{-1} \bar{b}_{-1}) c_0^+
			\\ &
			+ \I P'_{\mu}(k) \, (\alpha^{\mu}_{-1} \bar{b}_{-1} - b_{-1} \bar{\alpha}^{\mu}_{-1}) c_0^-c_0^+
			+ \I Q'_{\mu}(k) \, (\alpha^{\mu}_{-1} \bar{c}_{-1} + c_{-1} \bar{\alpha}^{\mu}_{-1})
			\\ &
			+ R'_{\mu\nu}(k) \, (\alpha^{\mu}_{-1} \bar{\alpha}^{\nu}_{-1} + \alpha^{\nu}_{-1} \bar{\alpha}^{\mu}_{-1}) c_0^+
			\\ &
			+ S'_{\mu\nu}(k) \, (\alpha^{\mu}_{-1} \bar{\alpha}^{\nu}_{-1} - \alpha^{\nu}_{-1} \bar{\alpha}^{\mu}_{-1}) c_0^-
			\Big) \ket{k, \downarrow\downarrow}.
	\end{aligned}
\end{equation}

\subsubsection{Gauge invariance and field redefinition}

\paragraph{Physical string field}

The gauge parameter reads:
\begin{equation}
	\ket{\Lambda_{1,1}^+(k)}
		= \Big(
			\chi^+(k) \, c_0^+ b_{-1} \bar b_{-1}
			+ \I \xi^+_\mu(k) \big( \alpha_{-1}^\mu \bar b_{-1} - b_{-1} \bar \alpha_{-1}^\mu \big)
			\Big) \ket{k, \downarrow\downarrow}.
\end{equation}

By acting with $Q_B$, we obtain the gauge transformations:
\begin{equation}
	\begin{aligned}
	\delta D^+(k)
		&
		= - \chi^+(k)
			+ \sqrt{\frac{\alpha'}{2}} \, \I k^{\mu} \xi^+_{\mu}(k),
	\\
	\delta E^+_{\mu}(k)
		&
		= - \frac{\alpha' k^2}{2} \, \xi^+_{\mu}(k)
			- \sqrt{\frac{\alpha'}{2}} \, \I k_{\mu} \chi^+(k),
	\qquad
	\delta A^+_{\mu}(k)
		= 0,
	\\
	\delta B_{\mu\nu}
		&
		= \frac{1}{2} \sqrt{\frac{\alpha'}{2}} \,
			\big(\I k_{\nu} \xi^+_{\mu}(k) - \I k_{\mu} \xi^+_{\nu}(k) \big).
		\end{aligned}
\end{equation}

The gauge parameter for the transformation of $\Lambda_{1,1}^+(k)$ is:
\begin{equation}
	\ket{\Omega_{1,1}^+(k)}
		= \omega^+(k) \, b_{-1} \bar b_{-1} \ket{k, \downarrow\downarrow},
\end{equation}
and we find:
\begin{equation}
	\delta \chi^+(k)
		= \frac{\alpha' k^2}{2} \, \omega^+(k),
	\qquad
	\delta \xi^+_\mu(k)
		= - \sqrt{\frac{\alpha'}{2}} \, \I k_\mu \omega^+(k).
\end{equation}
Hence, we can redefine the $\Lambda_{1,1}^+$ parameters as:
\begin{equation}
	\bar \chi^+(k)
		= \chi^+(k)
			- \sqrt{\frac{\alpha'}{2}} \, \I k^\mu \xi^+_\mu(k),
	\qquad
	\bar \xi^+_\mu(k)
		= \xi^+_\mu(k).
\end{equation}
This gives the new transformations for the $\Psi$ components:
\begin{equation}
	\begin{aligned}
	\delta D^+(k)
		&
		= - \bar \chi^+(k),
	\qquad
	\delta A^+_{\mu}(k)
		= 0,
	\\
	\delta E^+_{\mu}(k)
		&
		= \frac{\alpha'}{2} \, \I k^\mu \big(\I k_\nu \bar \xi^+_{\mu}(k) - \I k_\mu \bar \xi^+_{\nu}(k) \big)
			- \sqrt{\frac{\alpha'}{2}} \, \I k_{\mu} \bar \chi^+(k),
	\\
	\delta B_{\mu\nu}
		&
		= \frac{1}{2} \sqrt{\frac{\alpha'}{2}} \,
			\big(\I k_{\nu} \xi^+_{\mu}(k) - \I k_{\mu} \bar \xi^+_{\nu}(k) \big).
		\end{aligned}
\end{equation}
Note that the field strength of $\bar \xi^+_\mu(k)$ now appears in the transformations of both $E_\mu^+$ and $B_{\mu\nu}$.

The field $D^+$ is pure gauge, and we can make the following field redefinition to make $E^+$ gauge invariant~\cite{Asano:2012:ClosedStringField}:
\begin{equation}
    \bar E^+_{\mu}(k)
		= E^+_{\mu}(k)
			- \sqrt{\frac{\alpha'}{2}} \,
				(2 \I k^{\nu} B_{\mu\nu}(k) + \I k_{\mu} D^+(k)).
\end{equation}

\paragraph{Spurious string field}

The gauge parameter of the spurious field is
\begin{equation}
	\begin{aligned}
	\ket{\widetilde{\Lambda}_{1,1}^-(k)}
		= \Big(
			&
			\epsilon(k) \, c_0^- c_0^+ b_{-1} \bar{b}_{-1}
			+ \kappa(k) \, (b_{-1} \bar{c}_{-1} - c_{-1} \bar{b}_{-1})
			\\ &
			+ \I \theta_{\mu}(k) \, (\alpha^{\mu}_{-1} \bar{b}_{-1} + b_{-1} \bar{\alpha}^{\mu}_{-1}) c_0^+
			+ \I \lambda_{\mu}(k) \, (\alpha^{\mu}_{-1} \bar{b}_{-1} - b_{-1} \bar{\alpha}^{\mu}_{-1}) c_0^-
			\\ &
			+ \rho_{\mu\nu}(k) \, (\alpha^{\mu}_{-1}\bar{\alpha}^{\nu}_{-1}+\alpha^{\nu}_{-1}\bar{\alpha}^{\mu}_{-1})\big) \ket{k, \downarrow\downarrow}
	\end{aligned}
\end{equation}

So the gauge transformations we obtain by acting with $Q_B$ are
\begin{equation}
	\begin{aligned}
    \delta M'(k)
		&
		= \epsilon(k)
			+ \sqrt{\frac{\alpha'}{2}} \, \I k^{\mu} \lambda_{\mu}(k),
	\\
    \delta N'(k)
		&
		= \epsilon(k)
			- \frac{\alpha' k^2}{2} \, \kappa(k)
			- \sqrt{\frac{\alpha'}{2}} \, \I k^{\mu}\theta_{\mu}(k),
	\\
    \delta P'_{\mu}(k)
		&
		= \frac{\alpha' k^2}{2} \, \lambda_{\mu}(k)
			- \sqrt{\frac{\alpha'}{2}} \, \I k_{\mu} \epsilon(k),
	\\
    \delta Q'_{\mu}(k)
		&
		= \theta_{\mu}(k)
			- \lambda_{\mu}(k)
			- 2 \sqrt{\frac{\alpha'}{2}} \, \I k^{\mu} \rho_{\mu\nu}(k)
			- \sqrt{\frac{\alpha'}{2}} \, \I k_{\mu} \kappa(k),
	\\
    \delta R'_{\mu\nu}(k)
		&
		= \frac{\alpha' k^2}{2} \, \rho_{\mu\nu}(k)
			+ \frac{1}{2} \sqrt{\frac{\alpha'}{2}} \,
				\big( \I k_{\nu} \theta_{\mu}(k) + \I k_{\mu} \theta_{\nu}(k)\big),
	\\
    \delta S'_{\mu\nu}(k)
		&
		= \frac{1}{2} \sqrt{\frac{\alpha'}{2}} \,
			\big( \I k_{\nu} \lambda_{\mu}(k) - \I k_{\mu} \lambda_{\nu}(k) \big).
	\end{aligned}
\end{equation}

\subsubsection{Equation of motion}

\paragraph{Physical string field}

The equations of motion obtained from \eqref{eq:sft:eom} are:
\begin{equation}
	\begin{aligned}
	0
		&
		= \frac{\alpha' k^2}{2} \, D^+(k)
			+ \sqrt{\frac{\alpha'}{2}} \, \I k^{\mu} E^+_{\mu}(k),
	\\
	0
		&
		= \alpha' k^2 \, B_{\mu\nu}(k)
			+ \sqrt{\frac{\alpha'}{2}} \,
				\big(\I k_{\nu} E^+_{\mu}(k) - \I k_{\mu} E^+_{\nu}(k) \big),
	\\
	0
		&
		= \sqrt{2\alpha'} \, \I k^{\mu} B_{\mu\nu}(k)
			- \sqrt{\frac{\alpha'}{2}} \, \I k_{\nu} D^+(k)
			+ E^+_{\nu}(k)
			- A^+_{\nu}(k),
	\\
	0
		&
		= \frac{\alpha' k^2}{2} \, A^+_{\mu}(k),
	\qquad
	0
		= \sqrt{\frac{\alpha'}{2}} \, \I k^{\mu} A^+_{\mu}(k),
	\\
	0
		&
		= \sqrt{\frac{\alpha'}{2}} \,
			\big(\I k_{\nu} A^+_{\mu}(k) + \I k_{\mu} A^+_{\nu}(k) \big).
	\end{aligned}
\end{equation}

Let us prove that $A^+_\mu = 0$, implying that there is no additional propagating degree of freedom.
Writing $\mu = 0, 1, i$ with $i = 2, \ldots, d-1$ and $k^\mu = (k^0, k^1, k^i)$, the equation $k^2 A^+_\mu = 0$ implies $k^1 = \pm k^0$.
Then, $k^\mu A^+_\mu = 0$ yields $A^+_1 = \pm A^+_0$.
The last equation for $(\mu, \nu) = (0, i)$ gives $A^+_i = 0$, and $A^+_0 = A^+_1 = 0$ for $(\mu, \nu) = (0, 0)$.

After applying the field redefinition, the first equation becomes:
\begin{equation}
    0
		= \I k^{\mu} \bar{E}_{\mu}^+(k).
\end{equation}
We can obtain the usual equation of motion for the Kalb--Ramond field by multiplying the third equation with $k^\rho$ and anti-symmetrizing, before combining with the second equation:
\begin{equation}
    0
		= \alpha' \Big[
				k^2 B_{\mu\nu}
				+ k_{\mu} k^{\rho} B_{\nu\rho}
				- k_{\nu}k^{\rho}B_{\rho\mu}
				\Big]
			+ \sqrt{\frac{\alpha'}{2}} \,
				\big(\I k_{\nu} A^+_{\mu}(k) - \I k_{\mu} A^+_{\nu}(k) \big).
\end{equation}

\subsubsection{Action}

The action is:
\begin{align}
	S_{1,1}^+
		&
		\nonumber
		= \int \frac{d^d k}{(2\pi)^d} \bigg[
			- \frac{1}{4} \, M'(-k) \alpha' k^2 M'(k)
			- \sqrt{\frac{\alpha'}{2}} \, P'_{\mu}(-k) \I k^{\mu} M'(k)
			\\ & \hspace{2cm}
			\nonumber
			- \sqrt{\frac{\alpha'}{2}} \, S'_{\mu\nu}(-k)
			\big(\I k_{\nu} P'_{\mu}(k) - \I k_{\mu} P'_{\nu}(k) \big)
			\\ & \hspace{2cm}
			\nonumber
			- \frac{1}{2} \, P'^{\mu}(-k) P'_{\mu}(k)
			- \frac{1}{2} \, S'_{\mu\nu}(-k) \alpha' k^2 S'^{\mu\nu}(k)
			\bigg]
		\\ & \quad
		\nonumber
		+ \int \frac{d^d k}{(2\pi)^d} \bigg[
			- M'(-k) \bigg(
				\frac{\alpha' k^2}{2} \, D^+(k)
				+ \sqrt{\frac{\alpha'}{2}} \, \I k^{\mu} E^+_{\mu}(k)
				\bigg)
			\\ & \hspace{2.5cm}
			\nonumber
			+ \sqrt{\frac{\alpha'}{2}} \, N'(-k) \I k^{\mu} A^+_{\mu}(k)
			- \frac{1}{2} \, Q'^{\mu}(-k) \alpha' k^2 A^+_{\mu}(k)
			\\ & \hspace{2.5cm}
			\nonumber
			- P'^{\mu}(-k) \bigg(
				\sqrt{2\alpha'} \, \I k^{\mu} B_{\mu\nu}(k)
				- \sqrt{\frac{\alpha'}{2}} \, \I k_{\nu} D^+(k)
				\\ & \hspace{6cm}
				\nonumber
				+ E^+_{\nu}(k)
				- A^+_{\nu}(k)
				\bigg)
			\\ & \hspace{2.5cm}
			\nonumber
			+ S'^{\mu\nu}(-k) \bigg(
				\alpha' k^2 \, B_{\mu\nu}(k)
				+ \sqrt{\frac{\alpha'}{2}} \,
					\big(\I k_{\nu} E^+_{\mu}(k) - \I k_{\mu} E^+_{\nu}(k) \big)
				\bigg)
			\\ & \hspace{2.5cm}
			+ \sqrt{\frac{\alpha'}{2}} \, R'^{\mu\nu}(-k)
				\big(\I k_{\nu} A^+_{\mu}(k) + \I k_{\mu} A^+_{\nu}(k) \big)
				\bigg].
\end{align}

\subsection{Level (2, 0) -- non-level-matched massless fields}

\subsubsection{Field}

 \paragraph{Physical string field}

The string field reads:
\begin{equation}
	\begin{aligned}
    \ket{\Psi_{2,0}(k)}
		= \Big(
			&
			R(k) \, b_{-2} c_0^-
			+S(k) \, b_{-2} c_0^+
			+ X(k) \, b_{-1} c_{-1}
			\\ &
			+ \I Y_{\mu}(k) \, b_{-1} \alpha^{\mu}_{-1} c_0^+
			+ \I W_{\mu}(k) \, b_{-1} \alpha^{\mu}_{-1} c_0^-
			\\ &
			+ Z_{\mu\nu}(k) \, \alpha^{\mu}_{-1} \alpha^{\nu}_{-1}
			+ \I V_{\mu}(k) \, \alpha^{\mu}_{-2}
			\Big) \ket{k, \downarrow\downarrow}.
	\end{aligned}
\end{equation}

\paragraph{Spurious string field}

The spurious field is
\begin{equation}
\begin{aligned}
	\ket{\widetilde{\Psi}_{(2,0)}(k)}
		= \Big(
			&
			I(k) \, b_{-2} c_0^- c_0^+
			+ J(k) \, c_{-2}
			+ A(k) \, b_{-1} c_{-1} c_0^-
			+ B(k) \, b_{-1}c_{-1}c_0^+
			\\ &
			+ \I C_{\mu}(k) \, b_{-1} \alpha^{\mu}_{-1}c_0^- c_0^+
			+ \I D_{\mu}(k) \, c_{-1} \alpha^{\mu}_{-1}
			\\ &
			+ E_{\mu\nu}(k) \, \alpha^{\mu}_{-1} \alpha^{\nu}_{-1} c_0^-
			+ F_{\mu\nu}(k) \, \alpha^{\mu}_{-1} \alpha^{\nu}_{-1} c_0^+
			\\ &
			+ \I G_{\mu}(k) \, \alpha^{\mu}_{-2} c_0^+
			+ \I H_{\mu}(k) \, \alpha^{\mu}_{-2} c_0^-
			\Big) \ket{k, \downarrow\downarrow}.
\end{aligned}
\end{equation}

\subsubsection{Gauge invariance and field redefinition}

\paragraph{Physical string field}

The gauge parameter reads:
\begin{equation}
	\begin{aligned}
	\ket{\Lambda_{2,0}(k)}
		&
		= \Big(
			\alpha(k) \, b_{-2}
			+ \I \beta_{\mu}(k) \, b_{-1} \alpha_{-1}^{\mu}
			\Big) \ket{k, \downarrow\downarrow}.
	\end{aligned}
\end{equation}

The gauge transformations we obtain by acting with $Q_B$ are
\begin{equation}
	\begin{gathered}
	\delta R(k)
		= - 2 \alpha(k),
	\qquad
	\delta S(k)
		= - \frac{\alpha' k^2}{2} \, \alpha(k),
	\\
	\delta X(k)
		= - 3 \alpha(k)
			- \sqrt{\frac{\alpha'}{2}} \, \I k^{\mu} \beta_{\mu}(k),
	\\
	\delta Y_{\mu}(k)
		= - \frac{\alpha' k^2}{2} \, \beta_{\mu}(k),
	\qquad
	\delta W_{\mu}(k)
		= - 2 \beta_{\mu}(k),
	\\
	\delta Z_{\mu\nu}(k)
		= \frac{1}{2} \, \sqrt{\frac{\alpha'}{2}} \,
			\big( \I k_{\mu} \beta_{\nu}(k) + \I k_{\nu} \beta_{\mu}(k) \big),
	\qquad
	\delta V_{\mu}(k)
		= \beta_{\mu}(k).
	\end{gathered}
\end{equation}

We perform the following field redefinitions:
\begin{equation}
	\label{eq:exp:02-field-redef}
	\begin{gathered}
	\bar R(k)
		= R(k),
	\qquad
	\bar S(k)
		= S(k)
			- \frac{\alpha' k^2}{4} \, R(k),
	\\
	\bar X(k)
		= X(k)
			- \frac{3}{2} \, R(k)
			- \frac{1}{2} \sqrt{\frac{\alpha'}{2}} \, \I k^{\mu} W_{\mu}(k),
	\\
	\bar V_{\mu}(k)
		= V_{\mu}(k),
	\qquad
	\bar W_{\mu}(k)
		= W_{\mu}(k)
			+ 2 V_{\mu}(k),
	\\
	\bar Y_{\mu}(k)
		= Y_{\mu}(k)
			+ \frac{\alpha' k^2}{2} \, V_{\mu}(k),
	\\
	\bar Z_{\mu\nu}
		= Z_{\mu\nu}
			- \frac{1}{2} \, \sqrt{\frac{\alpha'}{2}} \,
				\big( \I k_{\mu} V_{\nu}(k) + \I k_{\nu} V_{\mu}(k) \big),
	\end{gathered}
\end{equation}
such that
\begin{equation}
	\begin{gathered}
	\delta \bar R(k)
		= - 2 \alpha(k),
	\qquad
	\delta V_{\mu}(k)
		= \beta_{\mu}(k),
	\qquad
	\delta \bar S(k)
		= 0,
	\\
	\delta \bar X(k)
		= 0,
	\qquad
	\delta \bar Y_{\mu}(k)
		= 0,
	\qquad
	\delta \bar W_{\mu}(k)
		= 0,
	\qquad
	\delta Z_{\mu\nu}(k)
		= 0.
	\end{gathered}
\end{equation}
This shows that $\bar R$ and $\bar V_\mu$ are pure gauge fields.

\paragraph{Spurious string field}

The gauge parameter of the spurious field is
\begin{equation}
\begin{aligned}
\ket{\widetilde{\Lambda}_{2,0}}
	= \Big(
		&
		\gamma(k) \, b_{-2} c_0^-
		+ \delta(k) \, b_{-2}c_0^+
		\\ &
		+ \epsilon(k) \, b_{-1} c_{-1}
		+ \I \kappa_{\mu}(k) \, b_{-1} \alpha^{\mu}_{-1} c_0^+
		+ \I \theta_{\mu}(k) \, b_{-1} \alpha^{\mu}_{-1} c_0^-
		\\ &
		+ \lambda_{\mu\nu}(k) \, \alpha^{\mu}_{-1} \alpha^{\nu}_{-1}
		+ \I \pi_{\mu}(k) \, \alpha^{\mu}_{-2}
		\Big) \ket{k,\downarrow\downarrow},
\end{aligned}
\end{equation}
and the gauge transformations we obtain by acting with $Q_B$ are
\begin{equation}
\begin{gathered}
	\delta I(k)
		= - 2 \delta(k)
			+ \frac{\alpha' k^2}{2} \, \gamma(k),
	\qquad
	\delta J(k)
		= 2 \delta(k)
			+ 2 \gamma(k)
			+ 3 \epsilon(k),
	\\
	\delta A(k)
		= 2 \epsilon(k)
			- \sqrt{\frac{\alpha'}{2}} \, \I k^{\mu} \theta_{\mu}(k)
			- 3 \gamma(k),
	\\
	\delta B(k)
		= \frac{\alpha' k^2}{2} \, \epsilon(k)
			- \sqrt{\frac{\alpha'}{2}} \, \I k^{\mu} \kappa_{\mu}(k)
			- 3 \delta(k),
	\\
	\delta C_{\mu}(k)
		= - 2 \kappa_{\mu}(k)
			+ \frac{\alpha' k^2}{2} \, \theta_{\mu}(k),
	\\
	\delta D_{\mu}(k)
		= \kappa_{\mu}(k)
			+ \theta_{\mu}(k)
			+ 2\pi_{\mu}(k)
			- \sqrt{2\alpha'} \, \I k^{\nu} \lambda_{\mu\nu}(k)
			- \sqrt{\frac{\alpha'}{2}} \, \I k_{\mu} \epsilon(k),
	\\
	\delta G_{\mu}(k)
		= \frac{\alpha' k^2}{2} \, \pi_{\mu}(k)
			+ \kappa_{\mu}(k),
	\qquad
	\delta H_{\mu}(k)
		= 2 \pi_{\mu}(k)
			+ \theta_{\mu}(k),
	\\
	\delta E_{\mu\nu}(k)
		= 2 \lambda_{\mu\nu}(k)
			+ \sqrt{\frac{\alpha'}{2}} \, \I k_{\mu} \theta_{\nu}(k),
	\\
	\delta F_{\mu\nu}(k)
		= \frac{\alpha' k^2}{2} \, \lambda_{\mu\nu}(k)
			+ \sqrt{\frac{\alpha'}{2}} \, \I k_{\mu}\kappa_{\nu}(k).
\end{gathered}
\end{equation}

\subsubsection{Equation of motion}

\paragraph{Physical string field}

The equations of motion are:
\begin{equation}
	\begin{gathered}
		0
			= -2 S(k)
				+ \frac{\alpha' k^2}{2} \, R(k),
		\qquad
		0
			= 2 X(k)
				- 3 R(k)
				- \sqrt{\frac{\alpha'}{2}} \, \I k^{\mu} W_{\mu}(k),
		\\
		0
			= Y_{\mu}(k)
				+ \frac{\alpha' k^2}{2} \, V_{\mu}(k),
		\qquad
		0
			= 2 S(k)
				+ 2 R(k)
				+ 3 X(k),
		\\
		0
			= - 2 Y_{\mu}(k)
				+ \frac{\alpha' k^2}{2} \, W_{\mu}(k),
		\qquad
		0
			= 2 V_{\mu}(k)
				+ W_{\mu}(k),
		\\
		0
			= Y_{\mu}(k)
				+ \frac{\alpha' k^2}{2} \, V_{\mu}(k),
		\qquad
		0
			= 2 S(k)
				+ 2 R(k)
				+ 3 X(k),
		\\
		0
			= 2 Z_{\mu\nu}(k)
				+ \frac{1}{2} \sqrt{\frac{\alpha'}{2}} \,
					\big( \I k_{\nu} W_{\mu}(k) + \I k_{\mu} W_{\nu}(k) \big),
		\\
		0
			= - 3 S(k)
				+ \frac{\alpha' k^2}{2} \, X(k)
				- \sqrt{\frac{\alpha'}{2}} \, \I k^{\mu}Y_{\mu}(k)
		\\
		0
			= \frac{\alpha' k^2}{2} \, Z_{\mu\nu}(k)
				+ \frac{1}{2} \sqrt{\frac{\alpha'}{2}} \,
					\big( \I k_{\nu} Y_{\mu}(k) + \I k_{\mu} Y_{\nu}(k) \big),
		\\
		0
			= Y_{\mu}(k)
				+ W_{\mu}(k)
				+ 2 V_{\mu}(k)
				- \sqrt{\frac{\alpha'}{2}} \, \I k_{\mu}X(k)
				- 2 \sqrt{\frac{\alpha'}{2}} \, \I k^{\nu} Z_{\mu\nu}(k).
	\end{gathered}
\end{equation}

Setting $R = V_\mu = 0$ through a gauge transformation, the equations of motion simplify to:
\begin{equation}
	0
		= \bar S(k)
		= \bar X(k)
		= \bar Y_{\mu}(k)
		= \bar W_{\mu}(k)
		= \bar Z_{\mu\nu}(k),
\end{equation}
which shows that there are no physical degrees of freedom.

\paragraph{Spurious string field}

The equations of motion are:
\begin{equation}
\begin{gathered}
	0
		= 3 A(k)
			- 2 J(k)
			- 2 I(k),
	\qquad
	0
		= B(k)
			- \frac{\alpha' k^2}{4} \, A(k),
	\\
	0
		=
			2 H_{\mu}(k)
			-2 D_{\mu}(k)
			- C_{\mu}(k)
			+ \sqrt{\frac{\alpha'}{2}} \, \I k^{\nu} E_{\mu\nu}(k)
			+ \sqrt{\frac{\alpha'}{2}} \I k_{\mu} A(k),
	\\
	0
		= 2 \I F_{\mu \nu}(k),
	\qquad
	0
		= \frac{\alpha' k^2}{2} \, E_{\mu \nu}(k),
	\\
	0
		= \sqrt{\frac{\alpha'}{2}} \, k^{\mu} E_{\mu \nu}(k)
	\qquad
	0
		= \sqrt{\frac{\alpha'}{2}} \, k^{\mu} F_{\mu \nu}(k)
	\qquad
	0
		= \sqrt{\frac{\alpha'}{2}} \, k_{\mu} C_{\nu}(k),
	\\
	0
		= C_{\mu}(k)
			+ 2 G_{\mu}(k)
			- \frac{\alpha' k^2}{2} \, H_{\mu}(k),
	\qquad
	0
		=
			2 I(k)
			+ 3 B(k)
			- \frac{\alpha' k^2}{2} \, J(k),
	\\
	0
		=
			C_{\mu}(k)
			+ 2 G_{\mu}(k)
			- \frac{\alpha' k^2}{2} \, D_{\mu}(k)
			+ \sqrt{\frac{\alpha'}{2}} \, \I k^{\nu} F_{\mu\nu}(k)
			+ \sqrt{\frac{\alpha'}{2}} \, \I k_{\mu} B(k),
	\\
	0
		=
			3 I(k)
			- \sqrt{\frac{\alpha'}{2}} \, \I k^{\mu} C_{\mu}(k).
\end{gathered}
\end{equation}

\subsubsection{Action}

The action is
\begin{equation}
\begin{aligned}
	S_{2,0}
		= \int \frac{\dd^d k}{(2\pi)^d} \Bigg[
			&
			I(-k) \bigg(
				S(k)
				+ R(k)
				+ \frac{3}{2} \, X(k)
				\bigg)
			+ J(-k) \bigg(
				S(k)
				- \frac{\alpha' k^2}{4} \, R(k)
				\bigg)
			\\ &
			+ A(-k) \bigg(
				\frac{\alpha' k^2}{4} \, X(k)
				- \frac{1}{2} \sqrt{\frac{\alpha'}{2}} \, \I k^{\mu} Y_{\mu}(k)
				- \frac{3}{2} \, S(k)
				\bigg)
			\\ &
			+ B(-k) \bigg(
				- X(k)
				+ \frac{1}{2} \sqrt{\frac{\alpha'}{2}} \, \I k^{\mu} W_{\mu}(k)
				+ \frac{3}{2} \, R(k)
				\bigg)
			\\ &
			+ \frac{1}{2} \, C^{\mu}(-k) \bigg(
				Y_{\mu}(k)
				+ W_{\mu}(k)
				+ 2 V_{\mu}(k)
				\\ &
				\hspace{2.5cm}
				- \sqrt{\frac{\alpha'}{2}} \big(
					2 \I k^{\nu} Z_{\mu\nu}
					+ \I k_{\mu} X(k)
					\big)
				\bigg)
			\\ &
			+ D^{\mu}(-k) \bigg(
				Y_{\mu}(k)
				- \frac{\alpha' k^2}{4} \, W_{\mu}(k)
				\bigg)
			\\ &
			- \frac{1}{2} \, E^{\mu\nu}(-k) \bigg(
				\alpha' k^2 \, Z_{\mu\nu}(k)
				+ \sqrt{\frac{\alpha'}{2}} \big(
					\I k_{\mu} Y_{\nu}(k)
					+ \I k_{\nu} Y_{\mu}(k)
					\big)
				\bigg)
			\\ &
			+ F^{\mu\nu}(-k) \bigg(
				2 Z_{\mu\nu}(k)
				+ \frac{1}{2} \sqrt{\frac{\alpha'}{2}} \big(
					\I k_{\mu} W_{\nu}(k)
					+ \I k_{\nu} W_{\mu}(k)
					\big)
				\bigg)
			\\ &
			+ G^{\mu}(-k) \Big(
				2 V_{\mu}(k)
				+ W_{\mu}(k)
				\Big)
			- H^{\mu}(-k) \bigg(
				\frac{\alpha' k^2}{2} \, V_{\mu}(k)
				+ Y_{\mu}(k)
				\bigg)
			\Bigg].
\end{aligned}
\end{equation}

\section{Discussion}

In this paper, we have pushed further the study of closed bosonic string field theory without the level-matching condition, which has been constructed~\cite{Okawa:2022:ClosedStringField}.
We have studied the equations of motion for the physical fields $\Psi$ up to the massless level, and given the action for all levels except $(2, 0)$.
In particular, we have studied the even sector of the $(1, 1)$ level, which contains the Kalb--Ramond $B$ field.
The total action computed in this way provides the quadratic part of the low-energy effective theory without level-matching.

As was observed in~\cite{Okawa:2022:ClosedStringField} and discussed in \Cref{sec:level-exp:11-odd}, string theory without level-matching may not be equivalent to string theory with level-matching condition at the non-perturbative level, since it admits additional background solutions.
It would be interesting to investigate these new backgrounds with more details, especially, from the point of view of $T$-duality and double field theory.
A first step would be to make contact with the sigma model fields associated to the new components of the string field, following~\cite{Yang:2005:ClosedStringTachyon}.
Another possibility is to reproduce the analysis from~\cite{Sen:2021:MasslessRRSector} for the Ramond sector to get an interpretation of the additional fields.

As discussed in~\cite{Okawa:2022:ClosedStringField}, it is not clear how to interpret interactions without level-matching.
Indeed, off-shell amplitudes (and string vertices) become multi-valued for states which are not level-matched: while it is always possible to write string vertices which project each input on level-matched states, this does not seem very interesting.
This also requires investigating the ghost-dilaton theorem~\cite{Bergman:1995:DilatonTheoremClosed,Rahman:1996:VacuumVerticesGhostdilaton,Belopolsky:1996:WhoChangesString}, by understanding both which operator changes the string coupling (see \Cref{ft:dilaton}) and how to deal with the more general interactions.
The solution could arise again by analogy with the superstring:\footnotemark{}
\footnotetext{%
    We would like to thank Atakan Hilmi Fırat for this observation.
}%
super-SFT can also be formulated in terms of super-Riemann surfaces~\cite{Ohmori:2018:OpenSuperstringField,Takezaki:2019:OpenSuperstringField} (see~\cite{Wang:2022:EquivalenceSRSPCO} for a proof of the equivalence), and PCOs appear after bosonization of the $\beta\gamma$ system and act as projectors.
Hence, one can ask if there is a generalized notion of Riemann surface which can incorporate string fields without level-matching.
Another avenue is to study hyperbolic string field theory~\cite{Moosavian:2019:HyperbolicGeometryClosed-1,Moosavian:2019:HyperbolicGeometryClosed-2,Costello:2019:HyperbolicStringVertices,Firat:2021:HyperbolicThreeStringVertex} without level-matching to see how it is affected.
We hope to come back to these issues later.

\section*{Acknowledgments}

We are very grateful to Yuji Okawa for many discussions and sharing the draft from~\cite{Okawa:2022:ClosedStringField}.
We would also like to thank Atakan Hilmi Fırat, Thomas Mohaupt and Barton Zwiebach for comments on the draft and discussions.

HE has received funding from the European Union's Horizon 2020 research and innovation program under the Marie Skłodowska-Curie grant agreement No 891169.
This work of HE is supported by the National Science Foundation under Cooperative Agreement PHY-2019786 (The NSF AI Institute for Artificial Intelligence and Fundamental Interactions, \url{http://iaifi.org/}).
The work of MM has been supported by an EPSRC DTP International Doctoral Scholarship (project number 2271092).

\appendix

\section{Useful formulas}
\label{app:formulas}

We consider closed bosonic string theory in the critical dimension, $d = 26$, described by a $2d$ CFT made of $d = 26$ free scalar fields $X^\mu$ ($\mu = 0, \ldots, d - 1$) and a $bc$ ghost system (see~\cite{Erbin:2021:StringFieldTheory} for more details).
Except when necessary, we provide formulas only for the holomorphic sector $\partial X^\mu(z)$, $b(z)$ and $c(z)$, since the formulas involving $\bar \partial X^\mu(z)$, $\bar b(z)$ and $\bar c(z)$ for the anti-holomorphic sector follow by adding a bar.
\begin{equation}
    \I \partial X^\mu(z)
        = \sqrt{\frac{\alpha'}{2}} \sum_{n \in \Z} \frac{\alpha_n^\mu}{z^{n+1}},
    \qquad
    b(z)
        = \sum_{n \in \Z} \frac{b_n}{z^{n+2}},
    \qquad
    c(z)
        = \sum_{n \in \Z} \frac{c_n}{z^{n-1}}.
\end{equation}
The zero-modes of $\partial X^\mu(z)$ and $\bar \partial X^\mu(z)$ are related to the momentum operator
\begin{equation}
    \alpha_0^\mu
        = \bar \alpha_0^\mu
        := \sqrt{\frac{\alpha'}{2}} \, p^\mu.
\end{equation}
whose eigenvalue $k^\mu$ gives the momentum of the string center-of-mass.

The energy--momentum tensor is:
\begin{equation}
    T(z)
        := T^{\text{m}} + T^{\text{gh}}
        = \sum_{n \in \Z} \frac{L_n}{z^{n+2}},
    \qquad
    L_n
        := L_n^{\text{m}} + L_n^{\text{gh}},
\end{equation}
where
\begin{equation}
    T^{\text{m}}
        = - \frac{1}{\alpha'} \, \partial X \cdot \partial X,
    \qquad
    T^{\text{gh}}
        = - 2 \, b \partial c - \partial b \, c,
\end{equation}
and
\begin{equation}
    L_n^{\text{m}}
        = \frac{1}{2} \sum_{n \in \Z} \alpha_n \cdot \alpha_{m-n},
    \qquad
    L_n^{\text{gh}}
        = \sum_{n \in \Z} (m + n) b_{m - n} c_{n}.
\end{equation}
It is also useful to introduce the following combinations of zero-modes:
\begin{equation}
    L_0^\pm
        := L_0 \pm \bar L_0,
    \qquad
    b_0^\pm
        := b_0 \pm \bar b_0
    \qquad
    c_0^\pm
        := \frac{1}{2} \, (c_0 \pm \bar c_0).
\end{equation}

The BRST charge admits the following useful decomposition:
\begin{align}
    \label{eq:QB}
    Q_B
        &
        := c_0^+ L_0^+ - b_0^+ M^+
            + c_0^- L_0^- - b_0^- M^-
            + \widehat{Q}_B^+,
    \\
    \nonumber
    \widehat Q^\pm
        &
        := \sum_{n \neq 0} ( c_{-n} L_n^{\text{m}} \pm \bar c_{-n} \bar L_n^{\text{m}} )
            \\ & \hspace{1cm}
            - \frac{1}{2} \sum_{\substack{m, n \neq 0 \\ m + n \neq 0}}
                (m - n) ( c_{-m} c_{-n} b_{m+n} \pm \bar c_{-m} \bar c_{-n} \bar b_{m+n} ),
    \\
    M^\pm
        &
        := \sum_{n > 0} n ( c_{-n} c_n \pm \bar c_{-n} \bar c_n ),
    \\
    L_0^+
        &
        := N + \bar N
            + \frac{\alpha'}{2} \, p^2
            - 2,
    \\
    L_0^-
        &
        := N - \bar N,
    \\
    L_m^{\text{m}}
        &
        := \frac{1}{2} \sum_{n \neq 0,m} \alpha_n \cdot \alpha_{m-n}
            + \sqrt{\frac{\alpha'}{2}} \, p \cdot \alpha_m
    \qquad
    (m \neq 0),
\end{align}
where the level and number operators are defined as:
\begin{align}
    N
        &
        := N^{\text{m}} + N^{\text{b}} + N^{\text{c}},
    \\
    N^{\text{m}}
        &
        := \sum_{n > 0} n N^{\text{m}}_n,
    \qquad
    N^{\text{m}}_n
        := \frac{1}{n} \, \alpha_{-n} \cdot \alpha_n,
    \\
    N^{\text{b}}
        &
        := \sum_{n > 0} n N^{\text{b}}_n,
    \qquad
    N^{\text{b}}_n
        := b_{-n} c_n
    \\
    N^{\text{c}}
        &
        := \sum_{n > 0} n N^{\text{c}}_n,
    \qquad
    N^{\text{c}}_n
        := c_{-n} b_n,
    \\
    N_{\text{gh}}
        &
        := \widehat N_{\text{gh}}
            + c_0 b_0 + \bar c_0 \bar b_0 + 2,
    \\
    \widehat N_{\text{gh}}
        &
        := \sum_{n > 0} (N^{\text{c}}_n - N^{\text{b}}_n)
\end{align}

The (anti)-commutation relations between the different modes appearing previously are:
\begin{align}
    [\alpha_m^\mu, \alpha_n^\nu]
        &
        = m \eta^{\mu\nu} \delta_{m+n},
    &
    \{ b_m, c_n \}
        &
        = \delta_{m+n},
    \\
    [L^{\text{m}}_n, \alpha_{-n}^\mu]
        &
        = n \alpha_{m-n}^\mu,
    \\
    [N^{\text{m}}_n, \alpha_{-n}^\mu]
        &
        = \delta_{m,n} \alpha_{-m}^\mu,
    &
    [N^{\text{m}}, \alpha_{-n}^\mu]
        &
        = m \alpha_{-m}^\mu,
    \\
    [N^{\text{b}}_n, b_{-n}]
        &
        = \delta_{m,n} b_{-m},
    &
    [N^{\text{c}}_n, c_{-n}]
        &
        = \delta_{m,n} c_{-m},
\end{align}

The string Hilbert space $\mc H$ is a Fock space built from the vacuum $\ket{k, \downarrow\downarrow}$ which is defined as:
\begin{gather}
    b_0^\pm \ket{k, \downarrow\downarrow}
        = 0,
    \qquad
    p^\mu \ket{k, \downarrow\downarrow}
        = k^\mu \ket{k, \downarrow\downarrow},
    \\
    n > 0:
    \qquad
    \begin{aligned}
        \alpha_n^\mu \ket{k, \downarrow\downarrow}
            &
            = 0,
        &
        L^{\text{m}}_n \ket{k, \downarrow\downarrow}
            &
            = 0,
        \\
        b_n \ket{k, \downarrow\downarrow}
            &
            = 0,
        &
        c_n \ket{k, \downarrow\downarrow}
            &
            = 0.
    \end{aligned}
\end{gather}
The previous definitions immediately imply:
\begin{equation}
    \begin{gathered}
    L_0^+ \ket{k, \downarrow\downarrow}
        = \left[ \frac{\alpha' k^2}{2} - 2 \right] \ket{k, \downarrow\downarrow},
    \\
    L_0^- \ket{k, \downarrow\downarrow}
        = \widehat Q^\pm \ket{k, \downarrow\downarrow}
        = M^\pm \ket{k, \downarrow\downarrow}
        = 0.
    \end{gathered}
\end{equation}

For all computations in \Cref{sec:level-exp} except at level $(2, 0)$, we need only the first terms in each expansion:
\begin{subequations}
\begin{align}
    \widehat Q_B^+
        &
        \sim c_{-1} L^{\text{m}}_1 + c_1 L^{\text{m}}_{-1}
            + \bar c_{-1} \bar L^{\text{m}}_1 + \bar c_1 \bar L^{\text{m}}_{-1},
    \\
    M^\pm
        &
        \sim c_{-1} c_1 \pm \bar c_{-1} \bar c_1,
    \\
    L^{\text{m}}_{\pm 1}
        &
        \sim \sqrt{\frac{\alpha'}{2}} \, k \cdot \alpha_{\pm 1},
    \\
    [L^{\text{m}}_1, \alpha_{-1}^\mu]
        &
        = \alpha_0^\mu
        = \sqrt{\frac{\alpha'}{2}} \, p^\mu,
    \\
    L^{\text{m}}_{-1} \ket{k, \downarrow\downarrow}
        &
        = \sqrt{\frac{\alpha'}{2}} \, k \cdot \alpha_{-1} \ket{k, \downarrow\downarrow},
    \\
    [\alpha_1^\mu, \alpha_{-1}^\nu]
        &= \eta^{\mu\nu},
    \qquad
    \{ b_1, c_{-1} \}
        = 1.
\end{align}
\end{subequations}

Euclidean and BPZ conjugations are respectively defined as:
\begin{equation}
    \begin{gathered}
    \adj{\Big( \lambda \, A_1 \cdots A_n \ket{0} \Big)}
        = \conj{\lambda} \, \bra{0} \adj{A_n} \cdots \adj{A_1},
    \\
    \adj{\phi_n}
        = \phi_{-n},
    \qquad
    \adj{\ket{k}}
        = \bra{k},
    \\
    \adj{\ket{k, \downarrow\downarrow}}
        = \bra{-k} \bar c_{-1} c_{-1}
        =: - \bra{k, \downarrow\downarrow},
    \end{gathered}
\end{equation}
and
\begin{equation}
    \label{eq:cft:bpz-conj}
    \begin{gathered}
    \bpz{\Big( \lambda \, A_1 \cdots A_n \ket{0} \Big)}
        = \lambda \, \bra{0} \bpz{A_1} \cdots \bpz{A_n},
    \\
    \bpz{\phi_n}
        = (-1)^h \phi_{-n},
    \qquad
    \bpz{\ket{k}}
        = \bra{-k},
    \\
    \bpz{\alpha_n}
        = - \alpha_{-n},
    \qquad
    \bpz{b_n}
        = b_{-n},
    \qquad
    \bpz{c_n}
        = - c_{-n},
    \\
    \bpz{\ket{k, \downarrow\downarrow}}
        = \bra{-k} c_{-1} \bar c_{-1}
        =: \bra{-k, \downarrow\downarrow},
    \end{gathered}
\end{equation}
where we use $I(z) = 1/z$ for the inversion, $\lambda \in \C$, and $\phi$ is any field with conformal weight $h$.
The BPZ operator is BPZ odd and self-adjoint:
\begin{equation}
    \bpz{Q_B}
        = - Q_B,
    \qquad
    \adj{Q_B}
        = Q_B.
\end{equation}

This allows to define the BPZ inner-product $\braket{\cdot, \cdot}$ which is normalized as:
\begin{equation}
    \label{eq:normalization-bpz}
    \bra{k, \downarrow\downarrow} c_0^- c_0^+ \ket{k', \downarrow\downarrow}
        = \bra{k,0} c_{-1} \bar c_{-1} c_0^- c_0^+ c_1 \bar c_1 \ket{k', 0}
        := \frac{1}{2} \, (2\pi)^d \delta^{(d)}(k - k')
\end{equation}
since
\begin{equation}
    \bra{-k, \downarrow\downarrow}
        := \ket{k, \downarrow\downarrow}^t
        = \bra{k} c_{-1} \bar c_{-1},
    \qquad
    c_0^- c_0^+
        = \frac{1}{2} \, c_0 \bar c_0.
\end{equation}
The inner-product has the following properties:
\begin{equation}
    \label{eq:cft:bpz-exch}
    \braket{A, B}
        = (-1)^{A B} \braket{B, A},
    \qquad
    \adj{\braket{A, B}}
        = \braket{\adj B, \adj A},
\end{equation}
where an operator in exponent indicates its Grassmann parity.

Finally, the worldsheet parity $\Omega$ exchanges the left- and right-moving modes:
\begin{equation}
    \Omega \alpha_n^\mu \Omega^{-1}
        = \bar \alpha_n^\mu,
    \qquad
    \Omega b_n \Omega^{-1}
        = \bar b_n,
    \qquad
    \Omega c_n \Omega^{-1}
        = \bar c_n,
\end{equation}

\printbibliography[heading=bibintoc]

\end{document}